\documentclass[sigconf,nonacm,10pt]{acmart}
\AtBeginDocument{%
  }

\usepackage[utf8]{inputenc} % allow utf-8 input
\usepackage[T1]{fontenc}    % use 8-bit T1 fonts
\usepackage{url}            % simple URL typesetting
\usepackage{booktabs}       % professional-quality tables
\usepackage{amsfonts}       % blackboard math symbols
\usepackage{nicefrac}       % compact symbols for 1/2, etc.
\usepackage{microtype}      % microtypography
\usepackage{lipsum}
\usepackage{graphicx} % graphics
\usepackage{subcaption}
\usepackage{caption}
\usepackage{tabularx}
\usepackage{multirow}
\usepackage{array}
\usepackage{ragged2e}
\newcolumntype{Y}{>{\centering\arraybackslash}X}

\begin{document}

%% Title
\title{Mesh Augmentation of LoRaWAN-based IoT Networks}
%%%% Cite as
%%%% Update your official citation here when published 
%\thanks{\textit{\underline{Citation}}: 
%\textbf{Authors. Title. Pages.... DOI:000000/11111.}} 

%% Authors and affiliations (ACM acmart style)
\settopmatter{authorsperrow=4}
\author{Ram Ramanathan}
\authornote{Work done while at goTenna Inc}
\email{ram@nlyticallc.com}
\affiliation{%
  \institution{NLytica LLC}
  \state{MA}
  \country{USA}
}

\author{Dmitrii Dugaev}
\email{dmitrii@gotenna.com}
\affiliation{%
  \institution{goTenna Inc.}
  \state{NY}
  \country{USA}
}

\author{Liang Tang}
\authornotemark[1]
\email{liangtanee@gmail.com}
\affiliation{%
  \institution{Meta Inc.}
  \state{NJ}
  \country{USA}
}

\author{Warren Ramanathan}
\authornotemark[1]
\email{wramanathan@gmail.com}
\affiliation{%
  \institution{Raman Labs}
  \state{NJ}
  \country{USA}
}

\begin{abstract}
LoRaWAN is a leading standard and technology for low-power, long-range Internet-of-Things (IoT) communications. However, its single-hop architecture results in limited range and excessive power consumption for end devices, especially when deployed in large, remote and RF-challenged environments. Existing solutions are either incompatible with LoRaWAN, or limit relaying to a single hop. We present LIMA, a protocol for transparently augmenting an existing or new LoRaWAN deployment with a mesh network of LIMA routers. LIMA increases the effective coverage range well beyond the maximum LoRa range via multi-hopping, and significantly reduces the energy consumed by end-devices. LIMA requires no changes to the end-device, the servers or the LoRaWAN standard. LIMA builds routes using path reversal, tunnels LoRaWAN messages over LIMA, provides transparent extension of the existing Adaptive Data Rate (ADR), and suppresses duplicate forwarding. We describe the construction of a prototype LIMA router, and evaluation at scale using ns-3 simulations. Simulations show that LIMA improves the delivery rate, scalability, end-device energy consumption and latency by up to 5x, 8x, 12.6x and 2.3x respectively. Table-top and outdoor testing with the LIMA prototype router confirm that LIMA can provide multi-hop range extension and energy gains transparently within LoRaWAN.
\end{abstract}

% Keywords. The author(s) should pick words that accurately describe

\maketitle

\section{Introduction}

There has been a rapid acceleration in the deployment of Low-Power Wide Area Networks (LPWANs) over the last decade. LoRaWAN~\cite{loraallianceRP002104LoRaWAN} is a prominent LPWAN standard that is particularly attractive due to its ability to operate in license-free bands, be deployable as a private as well as a public network, and provide indoor penetration. In a LoRaWAN network, end-devices such as sensors send information using LoRa, a long-range communication technology based on \textit{chirp spread spectrum (CSS)} to servers via one or more gateways. LoRaWAN use cases include water and gas metering, smart cities, precision agriculture, environmental monitoring, logistics, a set that continues to grow.

Despite its success, however, LoRaWAN has two key problems that limit its market penetration in some scenarios. First, the end-devices (sensors) need to be within LoRa range of a gateway. This is a problem in remote off-grid deployments 
%%%%+++%%%
such as environmental monitoring in the arctic, or fire monitoring in wildland forests. It is also an issue 
%%%%+++%%%
and in RF-challenged settings such as industrial IoT, and hard-to-reach water meters. Second, long range in LoRa comes at the expense of high energy consumption. The longer the distance to the gateway, the higher the spreading factor required by LoRa, and the greater the energy expended by the end-device. In fact, the energy expenditure per useful bit can increase by a factor of $>$20x as the distance increases from 1 km to 6 km~\cite{bouguera2018energy}. Longer range also results in reduced data rate. 
In many such cases, adding additional LoRaWAN gateways near the end device is not an option due to the lack of line power and/or Internet connectivity, prohibitive cost or installation difficulty. Satellite-based options such as Starlink are also prohibitively expensive and unwieldy for many such use cases.

% new format with footnote
\begin{figure*}[t]
\vspace{0.5cm}
\centering
%%%+++%%%
\includegraphics[width=0.9\linewidth]{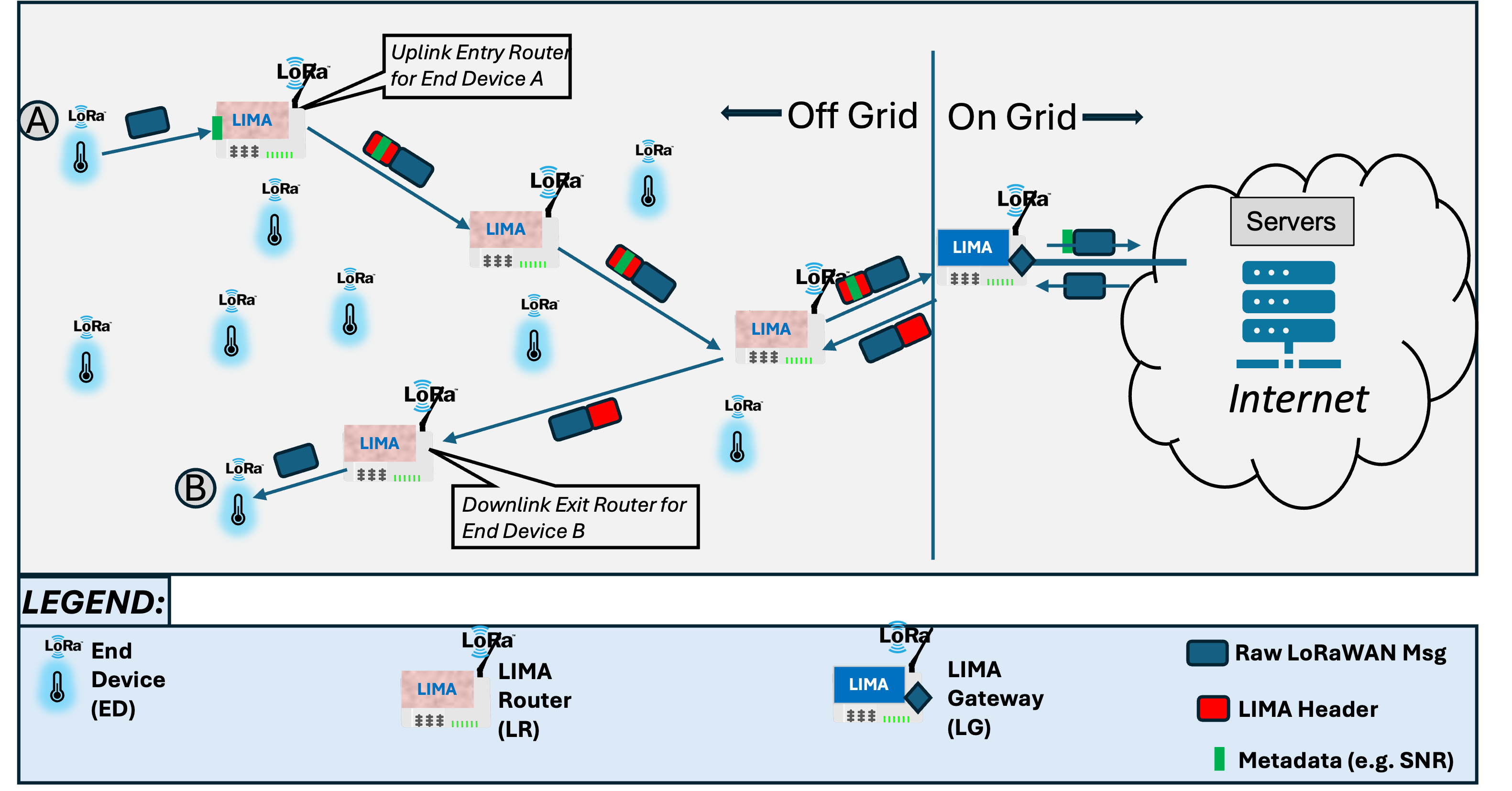}\par

{\footnotesize
\setlength{\parskip}{0pt}
\setlength{\baselineskip}{0.9\baselineskip}

\begin{minipage}{\linewidth}
\justifying
\noindent
LIMA mesh routers "dropped into" an existing LoRaWAN network provide transparent multi-hop uplink and downlink range extension and energy savings to end devices. Unlike existing solutions, LIMA does not require changes to LoRaWAN-compliant end devices or servers.
\end{minipage}\par
}

\vspace{-0.3cm}
\caption{LIMA operation.}
\label{fig:sys_overview}
\vspace{-0.4cm}
\end{figure*}

The current de facto solution to these problems is to add LoRaWAN gateways closer to the end devices. Unfortunately, in many use cases, including those mentioned above, this is not an option due the lack of line power and/or Internet connectivity, prohibitive cost or installation difficulty.
%While adding additional LoRgateways with cellular or other backhaul is possible, and is the de facto solution, it may be expensive, and may not practical in many cases as mentioned earlier.

In an attempt to address these issues, the LoRa Alliance has recently published the LoRaWAN Relay Specification~\cite{loraallianceRelaySpec}. A LoRaWAN Relay placed between end devices and the gateway can forward their messages to and from the gateway. However, it is only a partial solution since the standard only accommodates one relay, limiting the range extension. Further, the specification limits the number of end-devices that can use the relay, and requires changes to the end-devices using the relay, as well as to the LoRaWAN server. There is also a plethora of published work on mesh networking over LoRa, and a few products such as Meshtastic~\cite{Meshtastic} and NeoMesh~\cite{NeoMesh}, but these are traditional mesh networks that use LoRa as the physical layer, and are not compatible with LoRa\textit{WAN} -- for example, they cannot support critical LoRaWAN end-device functions such as over-the-air authentication, downlink RX window timing, end-device SF and power control etc. in a multi-hop manner, whereas LIMA can.

We present LoRaWAN IoT Mesh Augmentation (LIMA) -- a novel technique for seamlessly incorporating mesh networking within LoRaWAN. LIMA increases the effective coverage range to well beyond the maximum LoRa range via multi-hopping, significantly reduces the energy consumed by the end-devices, and reduces infrastructure costs by eliminating the need for additional backhauled gateways. Further, LIMA provides these benefits with no changes to the end-device, the servers, or the LoRaWAN standard, allowing self-organizing LIMA mesh routers to be simply "dropped in" to an existing LoRaWAN network. LIMA is a mesh routing protocol that works within the LoRaWAN standard in that the presence of the mesh network is transparent to both the server and the endpoint. 
 
Mesh augmentation of an existing or new LoRaWAN deployment using LIMA comprises of two platforms: a \textit{LIMA Router (LR)} and a \textit{LIMA Gateway (LG)}, both of which participate in running the LIMA protocol. An LG requires backhaul Internet connectivity as well as line power, whereas an LR does not require connectivity and can run on solar power or batteries. We have constructed the LG and LR using software modifications to a commercial LoRaWAN Gateway (see section \ref{sec:evaluation}), although other options including starting from scratch are possible. 
%%%%+++%%%
In particular, one could also use a commercial LoRaWAN Relay~\cite{loraallianceRelaySpec} product as a starting point; however, in this case, additional functionality pertaining to "waking up" a Relay using a WOR/WOR-Ack needs to be implemented. All of our work in this paper has been done assuming that an LR is based off of a Gateway, and is continuously awake.
%%%%+++%%%
LIMA includes several novel features that enable efficient routing and transparent multi-hop forwarding within a LoRaWAN network:

{\setlength{\leftmargini}{1.2em}
\begin{itemize}
\item Route creation via \textit{path reversal} techniques. Low-overhead Route Establishment Messages (REM) create reverse-path uplink routes, and uplink messages create downlink routes, enabling over-the-air activation of end-devices. 
\item Multi-hop uplink and downlink \textit{tunneling} of LoRaWAN messages using LoRaWAN-compatible header encapsulation.
\item Tunneled Adaptive Data Rate (ADR) that reacts to the SNR from the end-device (ED) to the \textit{nearest LIMA Router} rather than to the Gateway, thereby enabling significantly lower-energy operation.
\item Accommodating the end device Class A receive windows for multi-hop downlink forwarding, that is, managing the time of the last-hop LR to ED transmission to be in one of the two "awake" windows.
\item A distributed algorithm for selecting the \textit{designated edge router (DER)} from among many recipients of an ED message so as to efficiently transport an uplink message.
\item Creation and management of a Do Not Forward (DNoF) list that allows one to retain LoRaWAN operation for EDs that do not benefit from LIMA.
\end{itemize}
}

Although LIMA Routers are always awake, they do not need an internet backhaul, and so can be operated using solar power as evidenced by our simulations and real-world examples of solar-powered gateway deployments (e.g. see \cite{DryadSolarGateway}). Similar to gateways, LIMA Routers listen on all channels (region dependent), 
%All end-devices and LIMA Routers use the same channel, 
and thanks to the exceptional resilience of LoRa's CSS modulation to interference, it can receive simultaneously from multiple end devices. 

% An LR can be constructed in one of two broad ways:
% \begin{itemize}
% \item Using a standard LoRaWAN Gateway without backhaul. In this option, the LIMA protocol runs as a software modification on a commercial-off-the-shelf gateway. This LIMA Router (LR) does not need backhaul connectivity but likely requires solar power. No changes to the end devices or the LoRaWAN servers are required. A very small change to the actual LoRaWAN Gateway (the one with backhaul connectivity to the server) is required -- we call this upgraded gateway a \textit{LIMA Gateway} (LG).
% \item Using a LoRaWAN Relay compatible with the LoRAWAN Relay Spec~\cite{loraallianceRelaySpec}. In this option, the LIMA protocol runs a software modification on a commercial-of-the-shelf LoRaWAN Relay. Multiple such LIMA Routers based off of Relays form a mesh network that further extend the range, route around failures, and provide Adaptive Data Rate functions. In this option, we assume that all end-devices are "upgraded" to include functionality to interact with the Relays per specification.
% \end{itemize}

% \textit{As such the functionality of LIMA is agnostic to the platform -- gateway or end-device -- on which it runs. For simplicity of exposition, the reminder of this paper assumes that LIMA runs on a gateway, i.e., a LIMA Router (LR) is a modified LoRaWAN gateway. However, it should be emphasized that LIMA is platform-agnostic and can equally well run on a Relay.}

We have implemented a high-fidelity model of LIMA in the Network Simulator 3 (ns-3) tool and studied its performance over varying network size as well as traffic in a nominal remote monitoring deployment. In the region of parameters studied, LIMA outperforms LoRaWAN by a factor of up to 5x in delivery, 12.6x in energy consumption, 8x in coverage and 2.3x in latency. 

We have also developed a prototype LIMA Router and LIMA Gateway by modifiying the LoRaWAN software on a commercial LoRaWAN Gateway (the Seed Studio WM1302) and Raspberry Pi HAT add-on board. Using a Dragino temperature and humidity sensor and the The Things Network server, we performed both indoor and outdoor tests. We demonstrated multi-hop range extension as well as reduction in the energy consumed by the end device. We also showed that over-the-air join by end devices works over LIMA and that LIMA can co-exist within a LoRa-rich environment. 

LIMA can be used to augment existing or new LoRaWAN deployments in both private networks (e.g. precision agriculture, environmental monitoring etc.) and public IoT networks such as Helium~\cite{Helium} and Netmore~\cite{Netmore}. In private networks, users can benefit from higher coverage and longer sensor battery lifetimes without incurring the higher operating cost from additional gateways. In public networks, network service providers can fill gaps in coverage using LIMA to provide mesh connectivity at a fraction of the cost of deploying more gateways, each with its own backhaul.

%%%%+++%%%
The rest of the paper is organized as follows. We begin by discussing relevant background in section ~\ref{sec:Background}, and then provide a system overview in section~\ref{sec:system-overview}. Section~\ref{sec:LIMA-addressing-header-encapsulation} describes the packet format of the LIMA header. In sections~\ref{sec:route-establishment} and \ref{sec:message-forwarding} we describe how mesh routes are built and how they are used, respectively. Our evaluation of LIMA using ns-3 simulations and a prototype is described in section~\ref{sec:evaluation}. We conclude with a summary in section~\ref{sec:concluding-remarks}.
%%%%+++%%%

%\vspace{-0.1in}
\section{Background}
\label{sec:Background}

LIMA builds on two technologies -- LoRaWAN and mesh networking. In this section, we provide a background on LoRaWAN technology, as well as a review of the literature in the area of wireless mesh/multihop networking related to LoRa/LoRaWAN.
%\vspace{-0.2in}
\subsection{LoRaWAN Background}

LoRaWAN is a Low Power Wide Area Network (LPWAN) protocol that enables low-power Internet-of-Things (IoT) \textit{end devices} (EDs) to connect to network and application \textit{servers} on the Internet via one or more \textit{gateways}. LoRaWAN consists of two inter-related but distinct technologies: (1) LoRa ("Long Range"), a proprietary physical layer technology developed by Semtech Corporation for optimal long-range low-power operation; and (2) LoRaWAN, an open standard consisting mainly of a Medium Access (MAC) protocol that specifies the procedures for an end device, gateway and servers so that independent implementations can interoperate. The LoRaWAN specification is maintained by the LoRa Alliance~\cite{loraallianceTS001104LoRaWAN}. 
LoRaWAN devices are designed to operate autonomously for years. 
Other prominent LPWAN standards include NB-IoT~\cite{rohdeschwarzNarrowbandIoT}, LTE-M~\cite{LTE-M} and Sigfox~\cite{sigfoxSigfoxDevice}.
We briefly summarize LoRa and LoRaWAN below.

LoRa employs a proprietary variation of \textit{Chirp Spread Spectrum (CSS)} modulation~\cite{semtechLoRaLoRaWAN}, known for its resiliency under low SNR conditions.
Each information symbol in CSS is represented as a signal linearly swept across an entire bandwidth $B$: $(f_{0} - \frac{B}{2}, f_{0} + \frac{B}{2})$, where $f_{0}$ is carrier frequency, and $B$ is available channel bandwidth. The time duration of the symbol is varied according to an index called the \textit{Spreading Factor (SF)}. 
LoRa uses 6 Spreading Factors, ranging from SF7 to SF12. The lower the spreading factor, the higher the bitrate but lower the resiliency of the signal. 
Depending on the region, particular frequency set and available bandwidth, SF7 and SF12 can achieve speeds of around 27~\textit{kbps} and 0.98~\textit{kbps} respectively~\cite{semtechLoRaLoRaWAN}. LoRa can achieve up to 15 \textit{km} range in rural areas, and 5 \textit{km} in urban environments~\cite{semtechLoRaLoRaWAN}.

The LoRaWAN protocol is built on top of LoRa and operates over a "star-of-stars" topology, where a Network Server communicates with End Devices (EDs) (e.g. sensors/actuators) via one or more Gateways (GW) in the \textit{uplink} and \textit{downlink} directions. The GW is not power restricted and is equipped with a TCP/IP "backhaul" interface to communicate with the LoRaWAN Network Server. EDs are power-restricted devices which gather information in the field, and periodically transmit short burst data in the \textit{uplink} direction towards the GW. EDs have to be within 1-hop range of a GW. LoRaWAN networks are deployed in the unlicensed bands (e.g. 902–928 MHz in the US) and subject to duty cycle limitations.

EDs are divided into three classes: A, B, and C, depending on the power restrictions and the availability of \textit{downlink} resources~\cite{loraallianceTS001104LoRaWAN}. Only Class A implementation is mandatory, and an overwhelming majority of devices are Class A. We shall only consider Class A devices in this paper. Class A devices sleep for most of their time and open at most two receive windows RX1 and RX2 at pre-defined intervals after their uplink transmission. The LoRaWAN architecture places considerable responsibility on part of the servers to manage, authenticate and optimize the network. Notably, the network server dynamically assigns transmission parameters (SF, power etc) to an ED based on the SNR of its transmission to the gateway, thereby enabling low-power operation when the ED is close while not compromising connectivity when the ED is distant.

Given multiple SFs which LoRa-CSS may employ for communication, the LoRaWAN standard has incorporated \textit{Adaptive Data Rate (ADR)} feature~\cite{adr_analysis}. The ADR feature presents a mechanism that dynamically assigns a specific SF to a particular End-Device, based on the SNR at its uplink gateway. For instance, if an ED is located close to the gateway, the SNR value of the ED will be sensed as high at the gateway. This can trigger the ADR algorithm that will tell the ED to switch to a \textit{lower} SF to save on energy consumption and increase effective data rate. Due to substantial difference in energy consumption between SF7 and SF12, which can achieve \textit{20 times}~\cite{thethingsnetworkSpreadingFactor}, the ADR can be a powerful mechanism to significantly improve both the network lifetime and its capacity. For details on the ADR procedure and the formula for adjustments, we refer the reader to the LoRaWAN specification~\cite{loraallianceTS001104LoRaWAN}.

In 2022, the LoRa Alliance released the LoRaWAN Relay Specification~\cite{loraallianceRelaySpec} aimed at range extension using an appropriately placed relay device in scenarios where direct communication between end-devices and gateways is impeded by distance or physical barriers. We discuss this enhancement in the next section as part of the background on multihopping within LoRaWAN.

\subsection{Multihop Networking over LoRaWAN}

Prior work on mesh networking in the context of LoRa or LoRaWAN may be broadly classified into three groups: (1) mesh routing over LoRa; (2) relaying within LoRaWAN; (3) LoRa Repeaters. We discuss each of these below.

A popular approach taken by a number of researchers is to leverage the LoRa PHY layer of end-devices and build a routing solution on top of it. Most papers use the end-device as a router and run mesh routing between the devices -- a survey can be found in \cite{cotrim2020lorawan}. Examples include~\cite{hchenlee_lora_mesh,dlundell_lora_mesh} which present routing protocols built on top of the LoRa-enabled hardware. In contrast, authors in~\cite{dwijaksara_mhop_gw_lora} use gateways as a router and propose a multihop gateway-to-gateway (G2G) communication protocol for LoRaWAN. Additionally, researchers in~\cite{ahmar_smart_hop} present a multi-hop MAC layer protocol that reduces the number of end-devices operating at high SFs, and in~\cite{zhu_clustering, liao_ct_multihop, tian2023lorahop}, the authors attempt to build a coordinated multi-hop solution on top of LoRa using concurrent transmission.

While these works show the feasibility of building mesh networking on top of LoRa radios, the solutions are not compatible with existing LoRaWAN networks since they don't adhere to the LoRaWAN standard in terms of frame formats, sleep schedules, authentication procedures etc., and may require synchronization and scheduling not available in LoRaWAN. Moreover, these solutions suffer from elevated power consumption, since end-devices have to access uplink and downlink links asynchronously to support routing mechanisms on top.

Researchers in~\cite{ahmar_smart_hop} present a multi-hop MAC layer protocol that reduces the number of end-devices operating at high SFs. 
%while implementing low overhead routing in order to reduce data extraction times. T
The solution requires all nodes to have connectivity with the gateway and requires manual configuration of next hops. In~\cite{zhu_clustering, liao_ct_multihop, tian2023lorahop}, the authors attempt to build a coordinated multi-hop solution on top of LoRa using concurrent transmission. However, this is a complex approach, requiring synchronizing, slot scheduling and the ability to extract concurrently received packets.% Further, since these techniques require multiple nodes to relay, they drain the energy faster than traditional routing.  

The LoRaWAN Relay specification~\cite{loraallianceRelaySpec} introduces the concept of \textit{relay devices} which act as intermediaries between the end-devices and the network. The relay device is a standard end device that periodically listens on a dedicated channel for a Wake On Radio (WOR) message from an ED, which signals that the ED 
This message has a long enough preamble to allow the relay device to sleep, yet wake up when needed.
wants to transmit a message through the relay to the gateway. The specification requires all LoRaWAN devices to be modified to be able to use relaying. A relay device can support at most 16 EDs. Most importantly, the specification is limited to just one relay between an end-device and the gateway/server, and therefore, unlike a generalized mesh network, the range extension is limited to at most twice the original range (if that). 

Another approach similar to LoRaWAN relays is a \textit{LoRa repeater} (e.g. ~\cite{sisinni_lora_extender,khonrang2024development}). The repeaters present LoRa-PHY enabled devices that intercept LoRa signals from end-devices, and then retransmit them, thereby extending the effective range of the network. In other words, a LoRa repeater simply amplifies and retransmits LoRa signals. It works at the physical layer to extend the range of LoRa communication by receiving LoRa transmissions from a source device and then rebroadcasting them to extend the signal's reach.
The advantage of repeaters is that they can be LoRaWAN compatible without having to implement the Relay Specification. However, repeaters typically only work on uplink, cannot sleep and therefore may consume more power listening, and supporting multiple devices with a single repeater is challenging. They inherit the limitation of the LoRaWAN Relay in terms of a single additional hop.

A more recent "Gateway Mesh" product~\cite{rak-wireless-gateway-mesh} based on~\cite{chirpstack-gateway-mesh} has emerged with similarities to LIMA in terms of the encapsulation process. However, it appears that the construction of routes is based on manual configuration~\cite{chirpstack-GM-protocol}, which makes it unscalable.

LIMA is a full mesh routing solution that can provide range extension via multi-hop forwarding over several hops, and can route around failures. It rides on top of the broader LoRaWAN stack, encapsulates original packets, and ensures both multihop uplink and downlink connectivity among EDs and the GW. This enables LoRaWAN features such as ADR to work seamlessly, which increases the effective capacity of a network in a proximity of a LIMA-enabled router. LIMA routers can be deployed autonomously and self-organize into a mesh network. 
LIMA seamlessly augments LoRaWAN with full and transparent mesh networking capabilities without the need to modify end-devices or the server.  
Without being part of the LoRaWAN standard, LIMA can nonetheless provide LoRaWAN-compatible (or transparent) mesh networking without the need to modify end-devices or the server. 

Table~\ref{lima-vs-related} summarizes the attributes of each of the above class of approaches and LIMA. LIMA is the only solution that can seamlessly augment LoRaWAN with full and transparent mesh networking enabling scalable range extension, end-device energy reduction, support for downlink as well as uplink, and without modifications to the end-devices, server or the LoRaWAN standard.

\begin{table}[t]
\centering
\caption{LIMA vs related technologies.}
\vspace{-0.2cm}
\label{lima-vs-related}
\resizebox{\columnwidth}{!}{%
\begin{tabular}{|p{2.6cm}|c|c|c|c|c|}
\hline
 & \begin{tabular}[c]{@{}c@{}}\textbf{Scalable range-}\\ \textbf{extension}\\ \textbf{(multiple hops)}\end{tabular}
 & \begin{tabular}[c]{@{}c@{}}\textbf{Reduce ED}\\ \textbf{TX energy}\end{tabular}
 & \begin{tabular}[c]{@{}c@{}}\textbf{Downlink}\\ \textbf{as well as}\\ \textbf{uplink}\end{tabular}
 & \begin{tabular}[c]{@{}c@{}}\textbf{Use LoRaWAN}\\ \textbf{ED \& server}\\ \textbf{as-is}\end{tabular}
 & \begin{tabular}[c]{@{}c@{}}\textbf{LoRaWAN}\\ \textbf{standard}\\ \textbf{compatible}\end{tabular} \\
\hline
\begin{tabular}[c]{@{}l@{}}\textbf{LoRaWAN Relay}\\ \textbf{Specification}\end{tabular} & No  & Yes & Yes & No  & Yes \\
\hline
\textbf{LoRa Repeaters} & No & No & No & Yes & Yes \\
\hline
\begin{tabular}[c]{@{}l@{}}\textbf{LoRa-based}\\ \textbf{Mesh Network}\end{tabular} & Yes & N/A & No & N/A & No \\
\hline
\textbf{LIMA} & Yes & Yes & Yes & Yes & Yes \\
\hline
\end{tabular}%
}
\vspace{-0.4cm}
\end{table}

\section{System Overview}
\label{sec:system-overview}

As illustrated in Figure~\ref{fig:sys_overview} the LIMA system consists of a \textit{LIMA Router (LR)} and a LIMA Gateway (LG). Both LR and LG have a single LoRa interface for communication amongst themselves and with (unmodified) \textit{End Devices (ED)}. There could be several LRs placed in the area of deployment. An LG is modified LoRaWAN gateway and communicates with the \textit{Network Server (NS)} via an Internet interface. 

There may be one or more LGs, but typically just a few. 
LRs and LGs run the LIMA routing and tunneling protocol to form a self-organizing mesh network to convey messages over multiple “LIMA hops” from/to end devices. An LR that receives a raw LoRaWAN message from an ED is called the \textit{uplink entry LR} for that ED; similarly, an LR that sends a raw LoRaWAN message to an ED is called the \textit{downlink exit LR} for that ED. 

The goal of LIMA is not merely to create a stand-alone mesh protocol, but to \textit{augment LoRaWAN} with transparent mesh networking. In other words, LoRaWAN entities such as EDs and NSs must be unaware that a mesh network has been placed in between them. Thus, LIMA needs to go beyond simply tunneling LoRAWAN messages -- it needs to respect and work around the LoRaWAN standard and operational constraints. LIMA's LoRa messages should not affect EDs, and LIMA should be quiescent when multi-hop is not needed, and it should encapsulate metadata as well as messages. Key to achieving this is crafting a proper header structure that is efficient as well as LoRaWAN-compatible. In the rest of this section, we first present the LIMA header and summarize the operation.

% \begin{figure}[h]
% \centering
% \includegraphics[width=1.0\linewidth]{Figures/sys_overview_fig.png}
% \caption{LIMA Architecture and Components: LIMA Routers are placed to provide transparent multi-hop connectivity to end devices; the Gateway is updated to be a LIMA Gateway. No other changes to servers or end devices is needed.}% 
% \Description{Diagram showing LIMA Routers providing multi-hop connectivity between end devices and a LIMA Gateway.}
% %\vspace{-0.4cm}
% \label{fig:sys_overview}
% \end{figure}

\subsection{LIMA Header}
\label{sec:LIMA-addressing-header-encapsulation}
%%%%+++%%%
A LoRaWAN message generated by an end-device (ED) is received by a LIMA Router (LR), which encapsulates the message within a LIMA header and multi-hop forwards it over other LR until it reaches a LIMA Gateway (LG) where it is decapsulated and sent to the Server. In this section we discuss the details of the LIMA header and the constraints on its fields. 
%%%%+++%%%

In order to perform forwarding, a LIMA Node (LN), which could be an LR or an LG, needs to have a unique identifier/address. The LIMA protocol assigns addresses to LRs and LGs using an address space that is distinct from that used by LoRaWAN. Specifically, the LIMA address or \textit{identifier} of an LN is a 2-byte hash of some 8-byte hardware identifier that is guaranteed to be unique. For example, the unique hardware identifier could be the DevEUI of the LoRa chip for an LR, or the GatewayEUI for an LG. LIMA identifiers are used within the LIMA header to identify LIMA nodes and will be referred to as “LG/LR/LN id”. Non-LIMA entities such as EDs and Network/Application servers will not see these identifiers.

%%%++++%%% USE the width 1.0 for conference
\begin{figure*}[t]
\centering
\includegraphics[width=0.9\linewidth]{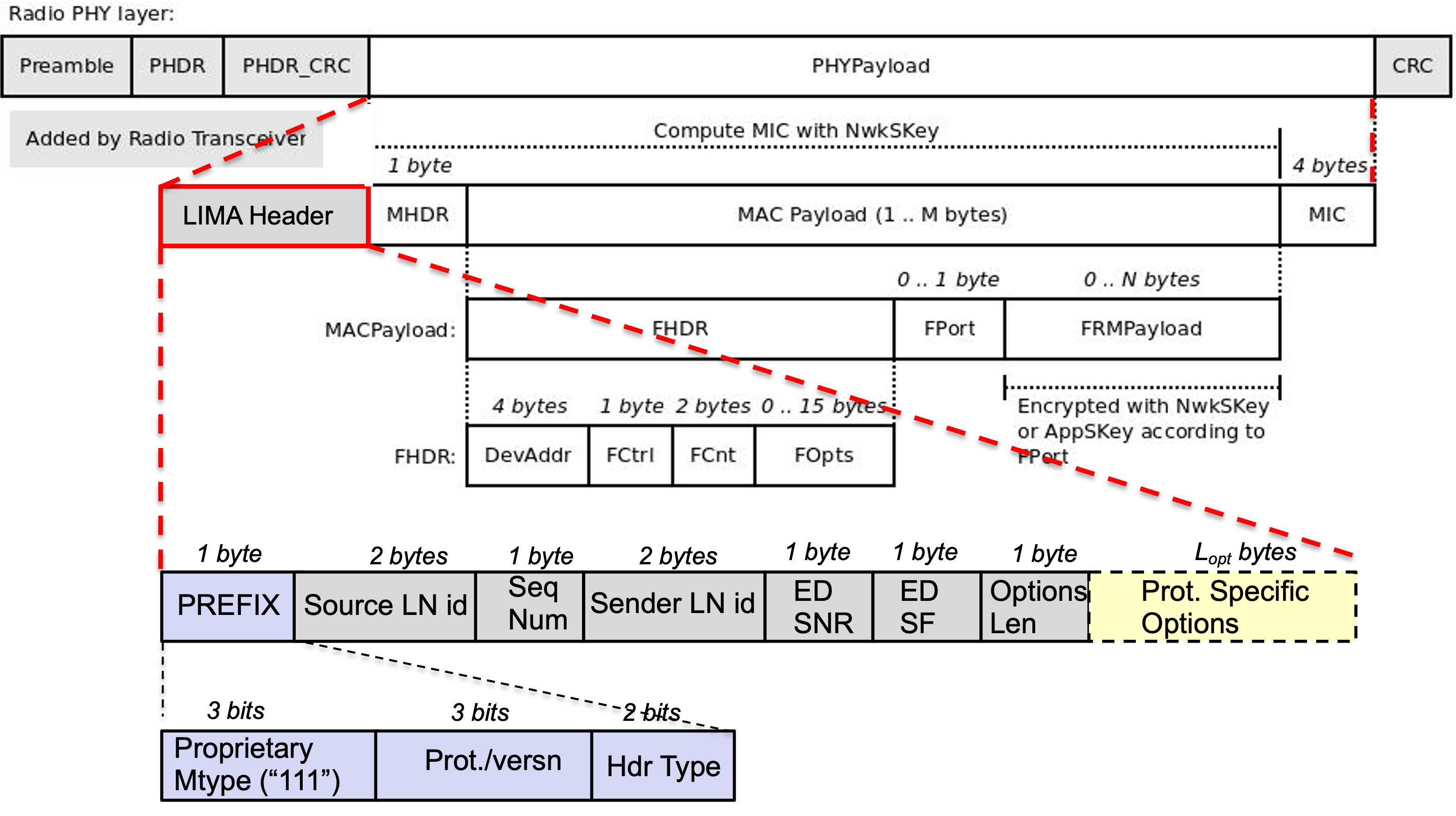}
\caption{The LIMA Header fields in the context of LoRaWAN message fields.}
\vspace{-0.4cm}
\label{fig:lima-header}
\end{figure*}

The LIMA header is used to transparently "tunnel" a LoRaWAN uplink or downlink message between the ED and the Server. Figure~\ref{fig:lima-header} shows the LIMA Header within the context of the LoRaWAN header~\cite{ttn-lorawan-format}. All fields shown within the expansion of the LIMA Header are new fields used only by LIMA, and all fields outside of that are LoRaWAN-specified fields. LIMA uses some of these fields but does not modify them. LIMA is at the MAC layer and above, and therefore "sees" only the PHYPayload from the LoRa radio. LIMA prepends a LIMA header to the PHYPayload (i.e., before the MHDR) as illustrated in  Figure~\ref{fig:lima-header}. The LIMA header fields can be grouped into three: a \textit{Prefix}, a set of fields for \textit{forwarding and metadata}, and \textit{protocol specific} fields. We now elaborate on these fields.

\paragraph{Prefix} The first sub-field of the Prefix (\textit{Proprietary MType}) is actually a LoRaWAN-related field to disambiguate LIMA messages from ED messages for both LIMA and the EDs. In LoRaWAN, the first three bits of the MHDR consists of the Frame Type (FType) which tells the recipient the kind of message is being sent (refer~\cite{loraallianceTS001104LoRaWAN} for a list of types). LoRaWAN specifies 111 to be a "proprietary" type, to be ignored by LoRaWAN entities. Accordingly, the first three bits of the Prefix, and hence the LIMA header as a whole, is set to 111\footnote{Without these bits, an ED will misinterpret the first three bits of an actual LIMA header as an Ftype.}. The \textit{Prot/version} tracks the version for backward compatibility purposes. The \textit{Hdr Type} indicates the type of header that follows: 0 for encapsulated uplink data, 1 for encasulated downlink data, 2 for Route Establishment Messages (REM).

\paragraph{Forwarding and Metadata} This group contains the fields shown following the Prefix in grey in Figure~\ref{fig:lima-header}. The \textit{source LN id} is the identifier of the originator of the message within the LIMA network, which also generates a \textit{sequence number} upon entry into the network. The sequence number is unique at the generator and the tuple (\textit{source LN id, seq num}) is globally unique and is used for detecting duplicates. The \textit{sender LN id} is the identifier of the LG or LR that transmitted this message. The ED SNR and SF are the signal-to-noise ratio and spreading factor respectively of the ED's message received by the uplink entry LR for the ED.

\paragraph{Protocol Specific Options} The options field, of length specified by the preceding \textit{options len} field (say L\textsubscript{opt} bytes) depends upon the \textit{header type} field in the Prefix. If this is a LoRaWAN data message (either uplink or downlink - type 0 or 1), then the options field contains the next hop target id (see section~\ref{sec:message-forwarding}), and hence L\textsubscript{opt} = 2 bytes. If this is an REM (type 2), then the options field contains the list of direct receivables (see section~\ref{sec:dnof-list-management}), and L\textsubscript{opt} is variable depending upon the size of this list, up to a maximum.

Apart from the above, LIMA uses fields as needed from the LoRaWAN header. Specifically, LIMA uses the DevAddr field in the MAC payload as a key to lookup downlink entries, except for JOIN-ACCEPT where it uses the DevEUI. We note that these fields are not encrypted on LoRaWAN and therefore can be read. While LoRaWAN has a Message Integrity Check (MIC) to detect changes, LIMA does not need to change any LoRaWAN fields.

%%%%+++%%%
The LIMA header is used to transparently "tunnel" a LoRaWAN uplink or downlink message between the ED and the Server. The original PHYPayload generated by the ED on uplink (or generated by the LG on behalf of the NS on downlink) is left intact following the above header. This includes the original MHDR, MAC Payload and the MIC. When “exiting” the LIMA network, the LIMA header is stripped off, yielding the original message in its entirety, providing end-to-end transparency.
%%%%+++%%%

LoRaWAN caps the length of the payload (PHYPayload in Figure~\ref{fig:lima-header} based on the Data Rate (DR) value and the region (see~\cite{loraallianceRP002104LoRaWAN}). 
%%%%+++%%%START
For example, in the U.S, the size of a  DR-0 (SF10) payload is capped at 11 bytes, whereas a DR-3 (SF7) payload is capped at 242 bytes. 
%%%%+++%%%END
Since LIMA adds additional fields, this will reduce the maximum number of payload bytes by 11 bytes. 
%%%%+++%%%START
Specifically, since the LIMA header without options is 9 bytes, and L\textsubscript{opt} = 2 bytes for data, the Application Payload cap is reduced due to LIMA by 11 bytes. This means that \textit{DR-0 cannot be supported in LIMA}. For DR-1, DR-2, DR-3 and DR-4, the effective payload sizes are 42 bytes, 114 bytes, 231 bytes and 231 bytes respectively, down from the original sizes of 53 bytes, 125 bytes, 242 bytes and 242 bytes respectively. 
This reduction is not an issue in practice since typical LoRaWAN payload sizes are small~\cite{fragkopoulos2023experimental}.

Finally, we note that the L\textsubscript{opt }is variable for REMs, but REMs do not contain data and hence this doesn’t affect payloads. However, this means that the bytes used for indicating direct receivables within REMs at a particular DR is limited to L\textsubscript{opt } $\le$ M - 9 bytes where M is the maximum allowed length for MAC Payload at that DR.

A LIMA implementation SHOULD check the incoming message and ensure that the MAC Payload (which includes the Application Payload -- see Figure~\ref{fig:lima-header}) is within the reduced limits. Specifically, the incoming MAC Payload should be at most M - 9 bytes, where M is the max MAC Payload size for the DR in use (see Table 20 in \cite{loraallianceRP002104LoRaWAN}). We note that in the absence of the "FOpts" field, the Application Payload (N) is equal to M-1. However, in the unlikely case of heavy use of MAC command piggybacking on the Fopts field, the payloads will be lower at low DRs.
%%%%+++%%%END

The LIMA header is prepended to the ED's LoRaWAN message by the uplink entry LR. The encapsulated message is multi-hop forwarded ("tunneled") over the mesh network of LRs to the LG. The LIMA header is stripped out by the LG before sending the original message to the network server. The reverse happens in the downlink direction -- the header is attached at the LG, tunneled through to the downlink exit LR which strips out the header and sends the vanilla LoRaWAN message to the ED. In the next section, we present an overview of LIMA's operation, with detailed descriptions in sections~\ref{sec:route-establishment} and \ref{sec:message-forwarding}.

\subsection{LIMA Operation: An Overview}

The core LIMA functionality is a mesh routing protocol that builds \textit{routing tables} to enable the forwarding mentioned above. The \textit{uplink routing table} at each LR contains the cost to reach each LG via each next hop LR, and is built using control packets called \textit{Routing Establishment Messages (REMs)} sent periodically by each LG. The uplink cost function is based on \textit{negative RSSI} (Received Signal Strength Indication) of the received REMs. 
The REMs are flooded throughout the LR network using standard flooding, and create uplink routings toward the source LG with an associated cost function that accumulates over each hop as the REMs percolate throughout the network. 
The \textit{downlink routing table} at each LR contains the cost to reach EDs, and is built by using uplink messages. The downlink cost function is based on hops. Specifically, both uplink and downlink route entries are populated using \textit{path reversal routing}, a technique wherein the route to a destination is the reverse of a path taken by a packet from that destination to the node. Downlink routes to EDs do not use REMs and are only built if an ED uplink message hops through the LR. An LR maintains multiple routes (a primary, and a configured number of backups) to LGs.

Uplink forwarding generally consists of following the least-cost route to an LG using the REM-built uplink routing tables. A separate technique is used, however, for the very first hop from the uplink entry LRs, namely the LRs that receive a raw LoRaWAN message. Since an ED cannot direct its message to a specific LR, multiple potential LRs may receive and forward it. To minimize this number and yet provide redundancy, a \textit{designated edge router} is distributively selected as the uplink entry LR for that ED. Uplink entry LRs also keep state and manage the timing regarding Class A EDs’ receive windows. These LRs buffer downlink messages to the EDs for which they are uplink entry points, and deliver them to the EDs when it opens up. 

Downlink forwarding consists of simply following the (reverse-path) downlink routing tables created during uplink message forwarding. 
The LG encapsulates the packet from the Network Server and forwards it to the next hop in its downlink routing table, and each LR does the same. 
The final LR/LG -- the downlink exit LR, which should be the same as the uplink entry LR mentioned above -- decapsulates the packet and broadcasts it to time with the ED reception windows based on the state stored during uplink reception. 

%The uplink routing table at an LR includes a cost, specifically the negative of the RSSI (Received Signal Strength Indication) that captures the total multi-hop cost of sending a message to each LG through each neighboring LR, i.e., each LR from which it received a REM. 
%Various cost functions are possible, from just using the hops to ones based on congestion, energy, etc. Currently, the cost function that we have used is simply the negative of the RSSI (Received Signal Strength Indication) of the most recent REM. Uplink forwarding follows the route that provides the least such cost to any LG.
%We recognize that the addition of RSSI values does not make sense semantically, but is chosen for its simplicity for now. The downlink routing table at an LR does not include a cost. The downlink path from an LG to an ED is simply the reverse of the most recent uplink path taken by a message from the ED to the LG.
 
LIMA routers do not forward messages from EDs that can efficiently reach the Gateway or LIMA Gateway directly. 
%at the lowest possible energy and/or spreading factor, as this will cause unnecessary transmissions. 
To accomplish this, each LR keeps a \textit{Do Not Forward (DNoF)} list of EDs whose messages it will not forward. To populate the DNoF, an LG includes a rolling list of \textit{direct receivables} – EDs that don’t require any assistance either for energy or for range – as part of its REM broadcasts. 
 
LoRaWAN includes an \textit{adaptive data rate (ADR)} mechanism wherein ther server adjusts the Spreading Factor (SF) and transmit power of an ED depending on SNR and SF received by the gateway. 
The ADR control is managed by the network server using the SNR from the ED to the gateway, which is measured by the gateway upon reception and sent as metadata to the network server.  
In LIMA, however, it is the SNR from the ED \textit{to the uplink entry LR} that is relevant. LIMA tunnels this SNR and SF to the server via the LG as part of its header. As a result, the server adjusts the SF and power to be optimal for the ED to LR link which requires much less energy for the ED. 
This is precisely the behavior we require, and one that reduces energy by exploiting the proximity of an ED to an LR.

Finally, we use the concept of a \textit{Transmission Profile (TP)}, which is a tuple of (SF, bandwidth, transmission power). A TP A is said to be higher than a TP B if A's SF and/or transmit power is higher than that of B (for the purposes of this work, the bandwidth is fixed at 125 KHz).
All LRs and LG(s) use the same predefined, shared standard transmission parameters, referred to as the \textit{Standard Transmission Profile (STP)}, which for this work we have set at SF7 and 125 KHz. 
%%%%+++%%%
Although the protocol does not dictate the exact STP, we recommend that the most energy-, throughput- and duty-cycle efficient combination for the specific region with sufficient range to provide a connected LIMA mesh network be used. Typically, this corresponds to SF7, 125 KHz and default power, and will be assumed unless specified otherwise. We assume that the LIMA router network is connected at this SF -- if not, the operator can add additional LIMA routers at appropriate locations to make the network connected, or use a higher STP.
%%%%+++%%%

%\input{LIMA-header}
\section{Route Establishment}
\label{sec:route-establishment}
%%%%+++%%%
A {\em route} is a sequence of nodes from a source to a destination such that each node in the sequence can directly reach the next node in the sequence. In a distributed routing mechanism such as LIMA, a route is implemented by having, at a given node, {\em forwarding or routing entries} for each destination. Each entry in such a {\em routing table} indicates the next-hop node for the destination of the message. Route establishment is the creation and maintenance of these forwarding entries, and is the subject of this section. 
%%%%+++%%%

The basic idea behind route establishment in LIMA, which we term \textit{path reversal routing} (PRR), is that if a node $X$ receives a packet that originated at a node $Y$ and transited through nodes $v_{1}$, $v_{2}$ ... $v_{k}$, in that order, then X has a route to Y by reversing the sequence of transit nodes, namely, $X$ $\rightarrow$ $v_{k}$ $\rightarrow$ $v_{k-1}$ $\rightarrow$ ... $\rightarrow$ $v_{1}$ $\rightarrow$ Y. LIMA creates PRR-based next-hop route entries based on information in the LIMA header. Due to its absence of periodic Hellos, link-state updates or route requests, PRR is simpler and more efficient than popular mesh routing protocols such as AODV~\cite{AODV} or OLSR~\cite{OLSR}. Variants of this basic idea have been used in \cite{BATMAN, SHARE, dusia2019vine}. PRR assumes link bidirectionality, which is valid in most LoRaWAN settings. We describe the procedure in more detail below.

\begin{figure*}[t]
\centering
%%%+++
\includegraphics[width=0.9\linewidth]{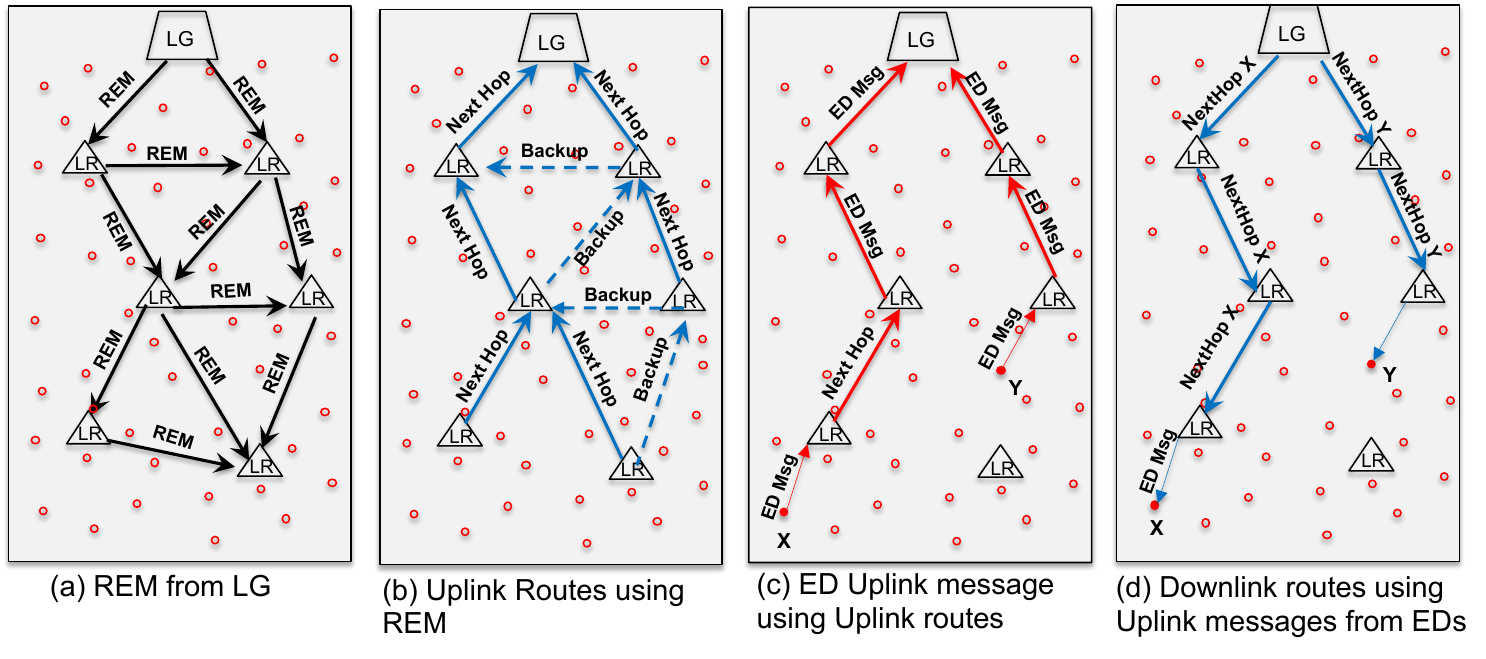}\par

{\footnotesize
\setlength{\parskip}{0pt}
\setlength{\baselineskip}{0.9\baselineskip}

\begin{minipage}{\linewidth}
\justifying
\noindent
(a) The LG broadcasts Route Establishment Messages (REM);
(b) Path-reversed uplink routes based on the REM are formed, both primary and backup;
(c) Uplink messages from example EDs X and Y are routed to the LG based on the uplink routes;
(d) and these enable path-reversed downlink routes to X and Y.
\end{minipage}\par
}

\vspace{-0.6em}
\caption{Route Establishment in LIMA between LR/LG supporting endpoints (shown as circles).}
\label{fig:route-establishment}
\end{figure*}

LIMA creates two routing tables: an {\em uplink routing table} that maintains routes toward each LIMA Gateway (LG), and a {\em downlink routing table} that maintains routes toward end-devices (EDs). The uplink routing table is constructed using periodic {\em Route Establishment Messages} (REMs) originated by LGs. The downlink routing table is constructed based on messages sent by the EDs to the LGs.  

Every LG periodically originates an REM. The REM uses the packet structure described in section~\ref{sec:LIMA-addressing-header-encapsulation}, with the header type field set to 2 to indicate that this is an REM. The field settings are as described in section~\ref{sec:LIMA-addressing-header-encapsulation}, with the protocol-specific option field containing the following:

{\setlength{\leftmargini}{1.2em}
\begin{itemize}
    \item \textit{Transmission Profile (SF, power, bandwidth) code}. The profile used to send the REM. LIMA will use a shared menu of possible SF, bandwidth, power combinations, each associated with a code. This field contains that code. A receiver can use the code to map back into the (SF, power, bandwidth) using the table.
    \item \textit{Cost from source}. Indicates the cost of reaching the source LG. Initialized at the LG to be zero. Each LR replaces this field by its own cost to the LG, computed as described below. If there are multiple LGs, this contains the least such cost. 
    \item \textit{(Optionally) Direct Receivables list}. A list of EDs from which the LG has directly received a message consistently. This enables LIMA to identify messages that it does not have to relay. More details are given in section~\ref{sec:dnof-list-management}
\end{itemize}
}

We note that each new REM has a sequence number unique to the generating source modulo wraparounds. The REM should preferably be sent on the downlink frequency channel that allows the highest duty cycle~\cite{loraallianceRP002104LoRaWAN} and preferably on SF7. The REM is broadcast using LoRa.

An LR that receives a REM sourced at S from another LR/LG (say R) executes the following procedure. First, if the sequence number of the REM is less than or equal to the sequence number tag of the routing entry for S as the destination, it discards the REM since the current entry is at least as fresh. Otherwise, it creates/updates the entry for destination S to have R as the next-hop node. It updates the cost to S with the cost received in the REM plus its own cost to R. LIMA allows any cost function to be used. For our LIMA simulations, we have used the negative of the RSSI. Finally, it tags this entry with the sequence number contained in the REM and retransmits the REM after modifying its \textit{sender-LN-id} and \textit{cost-from-source} fields appropriately. Specifically, it sets the \textit{sender-LN-id} to its own id, and increments the received \textit{cost-from-source} by the cost (-RSSI) from itself to the sender of the message. We note that wide fluctuations in environmental noise may affect the RSSI and make it non-monotonic; refinement and/or use of other cost functions such as the SNR is a subject for future research.

An LR creates an uplink routing entry for every additional neighbor from which a REM was received, regardless of whether this was a duplicate REM or not. These are "backup" entries and are used if the primary entry has expired. However, an LR only rebroadcasts the first received REM, after suitably updating the fields. 

The downlink routing table is also built in a similar manner using path reversal, but based on the uplink messages from an ED to an LG, which in turn are forwarded using the uplink routing table built as described above. Whereas the uplink entries are keyed by the LG address, the downlink entries are keyed by the ED address taken from the message. The LG and LRs create a new downlink entry toward the ED originating the message using its device EUI (DevEUI) if it is a JOIN-REQUEST, otherwise using the device address (DevAddr)\footnote{In LoRaWAN the server assigns the DevAddr as part of the JOIN-REPLY, hence for the JOIN-REQUEST we need to use the DevEUI)}. The next hop (downlink) is set to the ED address (DevEUI or DevAddr as the case may be) if the sender is an ED, and set to the LR address if not. A downlink route to an ED can only be formed after at least one uplink message from the ED. This is fine since an ED has to first send a JOIN-REQUEST for initialization, so there is guaranteed to be an uplink message preceding any downlink message for a given ED.

Figure~\ref{fig:route-establishment} summarizes the uplink and downlink route establishment using REM and uplink messages respectively.

%%%%+++%%%
We note that all route establishment is done at the LIMA routers (LRs) or the LIMA Gateways (LGs); the end-device and server(s) are completely unaware of this mechanism. We note that although the ED address is covered by the LoRaWAN MIC, it is still readable, and therefore can be used. An uplink message can be distinguished from a downlink message by the message-type field in the LIMA header.
%%%%+++%%%
\section{Message Forwarding}
\label{sec:message-forwarding}
LIMA multi-hop forwards uplink End Device (ED) messages to the Server and downlink Server messages to the ED using the uplink and downlink routing tables respectively (see section \ref{sec:route-establishment}). There are, however, several challenging questions that LIMA has to address in doing so: (1) how to forward in a manner transparent to the ED and Server, i.e, without modifying them; (2) on uplink, how to select a LIMA Router (LR) from among all that receive an ED message to forward, and how to forward only when needed, for example, not forwarding when an ED can reach the LG directly; (3) on downlink, since EDs have only two small receive windows, how do we manage the timing to send multi-hop over the LIMA network to ensure it arrives at the right time; (4) how to support Adaptive Data Rate (ADR). We discuss these under two main LIMA procedures for uplink and downlink data forwarding below.

\subsection{Uplink Message Forwarding}
\label{sec:uplink-message-forwarding}
%%%%+++%%%
There are three kinds of  uplink messages that can be received by a LIMA Node (LN): (1) from an ED directly to an LG, (2) from a LIMA Router (LR) to an LG, (3) from an ED to an LR, and  (4) from an LR to another LR. An LN determines which of these four cases applies as follows. An uplink message is from an ED if it has no LIMA header, and and the Frame type (FType) field in the LoRaWAN MAC header (MHDR) is 2 (“unconfirmed uplink”) or 4 (“confirmed uplink”). An  uplink message is from an LR if it has a LIMA header (identified by the first three bits of the PHYPayload being 111), AND the Header Type in the header is 0 (“00”). Thus, an LG/LR can decide which of the above four cases is true, and proceeds accordingly as below.
%%%%+++%%%

Uplink message forwarding through the LIMA mesh network to a Server via an LG may be thought of in three parts: (1) From the ED to multiple LRs, from amongst which one LR is selected to continue forwarding the message; (2) multi-hop through the LRs to the LG; and (3) from the LG to the Server. We describe the first in subsection \ref{sec:designated-edge-router} and the latter two in subsection \ref{sec:uplink-multi-hop-tunneling}.

A packet received at an LR is first checked to see if the ED address in the LoRaWAN header is in the LR’s Do-Not-Forward (DNoF) list (see section \ref{sec:dnof-list-management} for details) that contains a list of all EDs that are directly receivable. If so, it is silently discarded. The discussion below applies to those messages that are not in the DNoF and therefore need to be forwarded.

\subsubsection{Selecting the Designated Edge Router}
\label{sec:designated-edge-router}

Since there can be no target LR specified for a message transmitted by an unmodified ED, all LRs receiving the message could potentially forward the message. However, for efficiency reasons, one LR, termed the \textit{Designated Edge Router (DER)}, is selected to forward the message, as follows.

An LR initializes a Designated Map (DM) which contains a mapping between an ED and whether this LR is a DER for the ED or not. The DM is initially empty. Upon receiving a message from an ED, if the ED is not in the DM, then the LR forwards the packet using \textit{stagger relaying}. That is, it sets a timer randomly chosen between 0 and W (500 ms in our current implementation). If the LR overhears the same message being forwarded by another LR, then it cancels the timer and discards the message, otherwise it transmits the message when the timer expires. If it ends up transmitting, then it marks the DER status for the ED as True (this may change as described below). The LR continues to forward messages from the ED as long as its DM value is True.

At any point thereafter, if the LR overhears a LIMA message (i.e, one where the first three bits in the header are the LIMA Proprietary Indication (111)), it checks if the message is an uplink message using the header type. If so, it checks the \textit{ED SNR} field in the received header. Recall that an uplink entry LR places the SNR of the ED's message in this field. If this ED SNR is greater than the SNR of the most recently received message from the ED, then the DER status for the ED in the DM is marked False, that is, the LR perceives that another LR, namely the LR from which it received/overheard the message, is better suited than itself to be a DER, and hence “resigns”. The LR then marks its DER status for the ED as False and stops relaying packets from this ED. 

The above happens independently for each ED that is heard (processed) by the LR.

\begin{figure*}[h]
\centering
\includegraphics[width=0.9\linewidth]{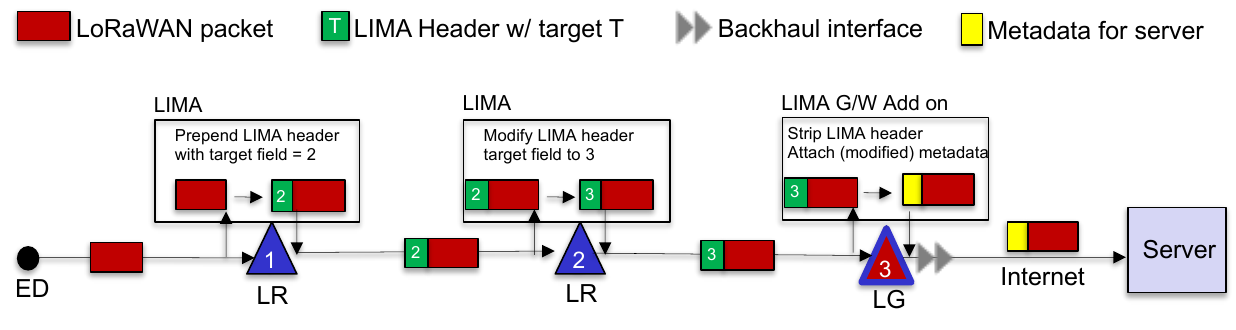}
\caption{Uplink Tunneling.}
\label{fig:uplink-tunneling}
\vspace{-0.4cm}
\end{figure*}

A message from an ED is forwarded by an LR if the ED's DER status is True, and dropped if the DER status is False. In a neighborhood of LRs, the LR that has the highest SNR from the ED remains as the DER after a couple of packet forwardings. 
%%%%+++%%%
If the multiple LRs that receive an EDs message are not all within range of each other, it is possible that two or more LRs remain DERs. This may result in redundant forwarding, but barring a very slight loss in efficiency, it is not an issue because LRs de-duplicate messages before forwarding (see section~\ref{sec:uplink-multi-hop-tunneling}). 
%%%%+++%%%
An entry is purged from the DM after a configured period which allows the DER status to be calculated afresh based on any topological changes that might have occurred.

The DER is not only the point of entry for uplink messages from a given ED but, by virtue of the downlink route establishment procedure, also the exit point for the downlink messages to it. To this end, to help manage the downlink timing to the ED, the DER calculates and stores the upcoming RX1 and RX2 times for the ED based on the LoRaWAN spec~\cite{loraallianceTS001104LoRaWAN}, along with other state information, such as the frequency channel on which it was received. The precise details or timing management are elaborated in section \ref{sec:end-device-receive-window-management}.

%%%%+++%%%
The use of a DER is an optional, efficiency-enhancing functionality. LIMA implementations can omit DER for simplicity reasons if required, without affecting the core functionality, but for a loss in bandwidth efficiency.
%%%%+++%%%

\subsubsection{Uplink Multi-hop Tunneling}
\label{sec:uplink-multi-hop-tunneling}
A Designated Edge Router (DER) that receives a message from an ED first checks if the ED is in the Do Not Forward (DNoF) list (details in section \ref{sec:dnof-list-management}) and if so discards it. If not, it prepends a LIMA header (refer section \ref{sec:LIMA-addressing-header-encapsulation}) to the message. The \textit{header type} is set to 0 indicating that is uplink data, and a new sequence number is generated. The source and sender-LN-id fields are both set to the LR id, and the SNR and SF fields is populated with the received SNR from the ED and the SF used respectively. The protocol-specific information field contains the next-hop LR/LG id, taken from the uplink routing table. 
%%%%+++%%%
If there are multiple entries, for example, one to each of several LGs, then the one with the least cost is chosen (ties broken randomly).
%%%%+++%%%
The message is then retransmitted. 

%%%%+++%%%
We note that since the first three bits of the message are set to the "111" (see section \ref{sec:LIMA-addressing-header-encapsulation}), which indicates "proprietary protocol" in LoRaWAN, any ED receiving this transmission will not further process it. Also, any LR that is not the next hop specified in the LIMA header also drops the packet. The intended next-hop replaces the sender-LN-id with its own id, sets the next hop target id per the uplink routing table, again picking the lowest-cost entry if there are multiple, and retransmits the message.
%%%%+++%%%

Figure~\ref{fig:uplink-tunneling} illustrates the "tunneling" of the LoRaWAN packet from an ED to Server through multiple LRs by attaching and detaching the LIMA header and changing the target id according to the uplink routing table.
At some point in this process, the target is an LG. An LG that receives the message strips out the LIMA header and retrieves the original LoRaWAN message. It then sends the message to the Server using standard LoRaWAN functionality except for a crucial difference -- the SNR in the metadata is taken from the SNR field of the LIMA header, as will be detailed in section \ref{sec:tunneled-adaptive-data-rate}, to enable Adaptive Data Rate. Note that this SNR is from the end-device to the uplink entry LR that was carried in the LIMA header across hops. 

No uplink message is transmitted more than once by an LR. Duplicate messages are identified and discarded using the same method that a server uses, in particular using the DevEUI, Fcnt (frame counter) fields within the LoRaWAN header and optionally the MIC which acts as a digital signature.

\subsubsection{Tunneled Adaptive Data Rate}
\label{sec:tunneled-adaptive-data-rate}
The LoRaWAN standard has an Adaptive Data Rate (ADR) feature which allows the Server to adjust an ED’s Data Rate (DR) value, namely the spreading factor and transmit power, so that it can most efficiently connect to the Gateway. If the ED is close to the gateway and has excess SNR compared to what is required, the server instructs the ED to lower its power and/or SF, and vice versa. The details of ADR include more features, and can be found in~\cite{loraallianceTS001104LoRaWAN}. 

%%%%+++%%%
This is done in three main steps per the standard:

{\setlength{\leftmargini}{1.2em}
\begin{enumerate}
    \item The Gateway passes the SNR of a received uplink packet in the metadata when sending the packet to the Network Server (NS).
    \item If needed, the Network Server computes new parameters for the ED to use, including the Data Rate (DR), transmit power, frequency and spreading factor (or an appropriate subset thereof), and sends an ADR request MAC command to the ED.
    \item The ED makes the change if possible and sends an ADR answer back to the NS.
\end{enumerate}
}
%%%%+++%%%

The challenge for LIMA is to adapt the ADR such that it adjusts the DR to be commensurate with the ED to LR (DER) link rather than the ED to Gateway/LG link. A further challenge is to do so without making changes to the ED or the NS. To accomplish this, LIMA has two sets of procedures, one at the uplink entry LR and the other at the LG.

At the uplink entry LR, the SNR of the message received from the ED is placed in the SNR field of the LIMA header, and the SF of the message received from the ED is placed in the SF field of the LIMA header. This message is transported to the LG using LIMA routing, and is placed in the metadata sent to the server. Thus, the SNR information of the link we want adapted, namely the ED-LR link, is "tunneled" through the LIMA mesh network to the NS.

%%%%+++%%%
We note that due to the Do-Not-Forward (DNoF) procedure (sections~\ref{sec:system-overview} and \ref{sec:dnof-list-management}), an ED message that is directly received with a Transmission Profile (TP) lower than the Standard Transmission Profile (STP) will not be forwarded via the LIMA network. For example, if the STP stipulates an SF of 8, EDs that can reach the LG directly with an SF of 8 or 7 will make it into the DNoF list. A direct ED-LG message with a TP at or lower than the STP is the vanilla LoRaWAN case and handled per the LoRaWAN specification.
%%%%+++%%%

When sending an ED message to the NS -- whether it was received directly or multi-hop via the LIMA network -- the LG records the maximum SNR value in the history log of size $h$, to account for SNR variations. For example, if $h$ = 3, and the five recent SNR values, temporally ordered, were 3.5, 7.7, 3.6, 7.6, 7.3, it uses 7.6 as the SNR to put in the metadata. It also includes the corresponding SF from the received LIMA header in the metadata.

%%%%+++%%%
The LG also overrides the “datarate” field of the metadata sent from the LG to the server with the information corresponding to the SNR. That is, whichever SNR was used to override, the corresponding SF from the same LIMA header is used also to override.  Since the LIMA header contains only the SF of the received packet, the SF is converted into the serialized DR string required by the metadata format and placed in the DR field. 
%%%%+++%%%

The network server uses the SNR and SF information from the metadata to calculate the SNR margin \cite{adr_analysis,adrSimpleDeterminationAlgorithm} in the same manner as it currently does, and computes the new SF and power if a change needs to be made. Note that the metadata used by the server actually contains the SNR/DR from the ED to the \textit{entry LR} and not the gateway, but the server is not aware of this.

Thus, the network server uses the SNR that the ED has to the best entry-point into the LIMA network. In turn, this makes the NS send an ADR-request to the ED \textit{to ramp down its SF to be sufficient to reach only the uplink entry LR, enabling considerable power savings}. By using only the most recent values of the SNR, we allow recovery if the LIMA router(s) were to become inoperational. For example, if the uplink entry LR were to go down, the LG will go back to sending the ED->LG SNR, which will then cause the ED to use the higher SF/power that is required to reach the LG.

%%%%+++%%%
Note that if two or more LRs pick up the ED message and forward (i.e there are two uplink entry LRs), the SNR used will be the maximum of the SNRs conveyed by each.
%%%%+++%%%

\subsubsection{Do Not Forward (DNoF) List Management}
\label{sec:dnof-list-management}

%%%%+++%%%
Every LR maintains a Do-Not-Forward (DNoF) list of EDs, which is initially empty. These are EDs that the LG deems can reach it directly (1 hop) with a sufficient signal strength if the ED transmits with a Transmission Profile (TP) at or lower than the Standard Transmission Profile (STP). Recall that a TP is lower than an STP if either the SF or power lower. For these EDs, multi-hopping through the LIMA network does not make sense. Upon receiving an uplink packet, an LR checks if the ED-address field is present in the DNoF list, and the entry is fresh enough. If so, then the message is silently discarded. Otherwise, the message is uplink forwarded as described in section \ref{sec:uplink-message-forwarding}. Note that this applies to messages directly received from an ED as well as through another LR.
%%%%+++%%%

The DNoF list is constructed at the LG using the direct receivable EDs, and then piggybacked on the REM packets sent by the LG. The direct receivables are tracked as follows. Upon successfully receiving an uplink packet, the LG checks if it was received from an ED, i.e. the first three bits of the PHY payload are NOT “111” (refer section \ref{sec:LIMA-addressing-header-encapsulation}). For such packets, it checks if the packet was sent with a TP at or below the STP. For example, if the STP uses SF7 and the incoming TP uses SF7, with the power and bandwidth being the same, then the check passes; if the TP has SF9, then it fails. If the check passes, then it adds the ED address to the direct receivables list, along with a timestamp that is used to age out old entries. 

When the time comes for the LG to send a REM, it selects a set of EDs that can fit within the REM based on the LoRaWAN payload size limitation for that frequency and SF. To select, the LG uses a circular buffer of ED addresses with a pointer moving over all of the EDs that are included in the current REM, and ending up pointing to the first un-included ED in the buffer. For the next REM, the procedure is repeated from where the pointer is left off. An ED is only included once in each REM. The direct receivables are placed in the protocol specific options field of the header.

Upon receiving a REM, an LR unpacks the list of direct receivables and populates its DNoF list accordingly. It updates the timestamp of existing entries and adds new entries. Entries older than a configured period are deleted. If there are multiple LGs, the LR may receive a different set of direct receivables from each LG. In this case, the DNoF represents the union of the direct receivables of all LGs. The suppression of uplink forwarding does not depend upon the originating LG of the DNoF item. The LR then broadcast forwards the REM as discussed in section \ref{sec:route-establishment}. Note that the direct receivable list is available to all LRs in the network, and so any LR can suppress the forwarding.

%%%%+++%%%
Note that right after ED initialization the DNoF list is empty and so all packets are forwarded. But very soon thereafter, the DNoF list is populated and only messages that cannot be directly received are forwarded.
%%%%+++%%%

Note that an ED that can reach the LG directly but with a high SF (e.g. SF12) is not included in the direct receivables (assuming STP < 12). Thus, it will not make it into the DNoF list. This will allow the ED’s message to be forwarded via LIMA as well, and the associated SNR to its uplink entry LR will be provided to the NS, and reduce its transmit power and/or SF via ADR.

%%%%+++%%%
The DNoF is an optional, efficiency enhancing feature. LIMA implementations can omit DNoF for simplicity purposes if required without affecting its core functionality.
%%%%+++%%%

\subsection{Downlink Message Forwarding}
\label{sec:downlink-message-forwarding}
A downlink packet for an ED received by an LG from the Network Server (NS) is forwarded based on the downlink routing table entry for the ED. The downlink routing table entry at the LG and each LR is created as was discussed in section \ref{sec:route-establishment}.
%%%%+++%%%
If the least-cost next-hop is the ED itself (direct link), then it is transmitted using the Spreading Factor (SF) in the NS packet’s metadata, as it would be done in the vanilla LoRaWAN. If the least-cost next-hop for the ED is an LR, it is forwarded using the STP (e.g. SF7), no matter what the NS intended it to be forwarded at -- i.e, the downlink SF is overridden by the LG to take advantage of the multi-hop. In this case, the message is tunneled using LRs to the ED in a manner similar to uplink multihop tunneling (section \ref{sec:uplink-multi-hop-tunneling}), except that it uses the downlink routing table entries, and does ED timing window management on the exit LR. We describe these procedures below.
%%%%+++%%%

\subsubsection{Downlink Multihop Tunnelling}
\label{sec:downlink-multihop-tunneling}
An LG receiving a message from the NS prepends a LIMA header (described in section \ref{sec:LIMA-addressing-header-encapsulation}) to the message. The \textit{header type} is set to 1 indicating that is is downlink data, and a new sequence number is generated and inserted. The source and sender LN id fields are set to the (own) LG id. The protocol specific information field contains the next-hop LR id, taken from the downlink routing table entry for the destination ED. We note that the lookup is based on the DevEUI if the message is a JOIN-ACCEPT and a DevAddr if not. This is because the DevAddr is not established during the initial JOIN phase. The message is then re-transmitted. 

%%%%+++%%%
We note that since the first three bits of the message are set to the "111" (refer section \ref{sec:LIMA-addressing-header-encapsulation}), which indicates "proprietary protocol" in LoRaWAN, any ED receiving this transmission will not further process it. Also, any LR that is not the target specified in the LIMA header also drops the packet. The target next-hop LR replaces the sender-LN-id with its own id, and sets the next hop LR id per the downlink routing table.
%%%%+++%%%

\begin{figure*}[t]
\centering
\includegraphics[width=0.9\linewidth]{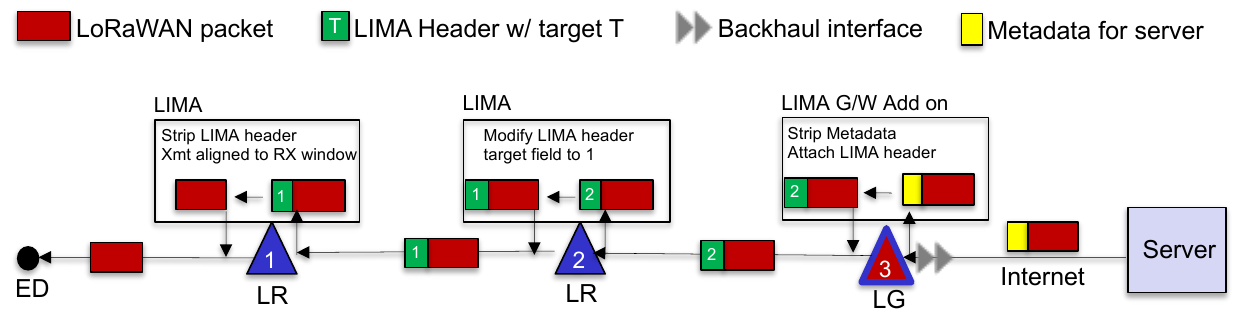}
\caption{Downlink Tunneling.}
\label{fig:downlink-tunneling}
\vspace{-0.4cm}
\end{figure*}

Figure \ref{fig:downlink-tunneling} illustrates the “tunneling” of the LoRaWAN packet from a Server to an ED through multiple LRs by attaching and detaching the LIMA header, and changing the target id as per the downlink routing table. An LR identifies duplicates using the (source-LN-id, sequence-number) fields and discards them.% For downlink messages, the source-LN-id is id of the LG that received the message from the server.

At some point in the sequence of multi-hop transmissions, the message reaches an LR where the next hop is the ED address, indicating the end of the tunneling process. The LR then strips out the LIMA header and retrieves the LoRaWAN-formatted message. However, in order to make sure the ED receives it, it needs to transmit it at a time when the ED is awake. This process is described in the next section.

\subsubsection{End-Device Receive Window Timing Management}
\label{sec:end-device-receive-window-management}
Message reception for Class A devices in LoRaWAN is only possible within two windows (RX1 and RX2) that are at a specific interval after an uplink transmission \cite{loraallianceTS001104LoRaWAN}. In vanilla LoRaWAN, the NS manages the timing such that the message is transmitted within one of those windows. In LIMA however, multi-hopping places a severe challenge to the timing constraint since the total time to reach the ED is highly variable.

To ensure transmission in one of these two windows, the ED's downlink exit LR (which also ought to be its uplink entry LR) keeps track of upcoming receive windows for the ED, buffers packets, and delivers the buffered packets in a FIFO manner. We describe this process in more detail below.

As mentioned in section \ref{sec:designated-edge-router}, the designated edge router (DER) for a given ED calculates and stores the upcoming RX1 and RX2 windows for the ED, along with other state information such as the frequency channel it was received on. Upon receiving any message, an LR checks if the packet is to be forwarded to an ED (i.e., the next hop in the downlink routing table is an ED, which implies this LR is the downlink exit LR). It then checks if it has RX timing and channel state for the ED. If so, it places it at the back of an “ED RX” queue for that ED. 

When the next window RX1 or RX2 occurs, the message from the front of the queue is taken and transmitted to the ED. We note that at this point the LIMA header has been stripped out and the message is the pure LoRaWAN message generated at the NS. The transmission parameters (frequency channel, data rate etc.) of this downlink transmission from the LR are a function of the uplink information it had stored earlier. The mapping between uplink channel and downlink channel is region dependent \cite{loraallianceRP002104LoRaWAN}. For example, in the EU the downlink frequency is the same as uplink whereas in the U.S it is the uplink channel index modulo 8 in the downlink set.% (note that in the U.S the uplink and downlink use disjoint channels). 

%%%%+++%%%
Note that this window could be corresponding to the uplink transmission for which the Server has responded, or it could be any future one. The LR simply takes the packet at the front of the queue, if any, when the window occurs and sends it to the ED. A packet that has stayed in the queue for longer than a configured period of time is discarded. Further, if the message was transmitted on an RX1 opportunity, then the timer for the RX2 opportunity is canceled since per LoRaWAN standard an ED only opens the RX2 if the RX1 is not used. 
%%%%+++%%%

It may be argued that if the multi-hop downlink message corresponding to an uplink message (e.g. an ACK) ``misses" the RX1 and RX2 windows, it may have to wait quite a long time for the next opportunity, which is after the next uplink message. While this is true, it is much more likely that the downlink message "catches" the RX1 or RX2 windows since the downlink LIMA messages are sent at the STP (SF7) whose delay even with several hops is less than the default 1 second budget for RX1. For example, with a 25 byte payload plus the LIMA and the LoRaWAN headers, the transmission delay per hop is about 97ms at SF7 125kHz BW. Given negligible propagation delay at the applicable distances and processing times in milliseconds, the hop budget for catching the RX1 is 8 hops and for catching the RX2 is 16 hops. In the rare case both windows are missed, then the message is stored for the next opportunity.

%%%%+++%%%
We note that, despite DER selection, multiple LRs may be uplink entry LRs for a given ED, and might have forwarded the packet in the uplink. However, a response downlink packet typically comes through a single LR (since each LR in the downlink direction is forced to make a choice between next hops and can use only one), and therefore should avoid collision at the ED. 
%%%%+++%%%

The LoRaWAN standard stipulates certain rules for RX2 such as SF and frequency. The LR must abide by these rules when using RX2 transmission window.

\section{LIMA Evaluation}
\label{sec:evaluation}
\subsection{Modeling and Simulation}
\label{sec:simulation}
This section describes the design, implementation, and evaluation of LIMA using the ns-3 simulation tool \cite{ns3}. The model was built on top of the existing LoRaWAN ns-3 module developed by D. Magrin et. al.~\cite{magrin2017performance}. The codebase of the LoRaWAN model \cite{lorawan-ns3-module} presents a standalone ns-3 plugin that contains a comprehensive implementation of the main LoRaWAN functionality that closely follows official LoRaWAN specifications.
%%%%+++%%%
Since most of the LoRaWAN features are reflected in the model, it provides a solid foundation that our LIMA model can extend for modeling both the LIMA Gateway (LG) and the LIMA Router (LR).
%%%%+++%%%

\subsubsection{Simulation Scenario}

\begin{figure}[ht]
\centering
\includegraphics[width=0.7\linewidth]{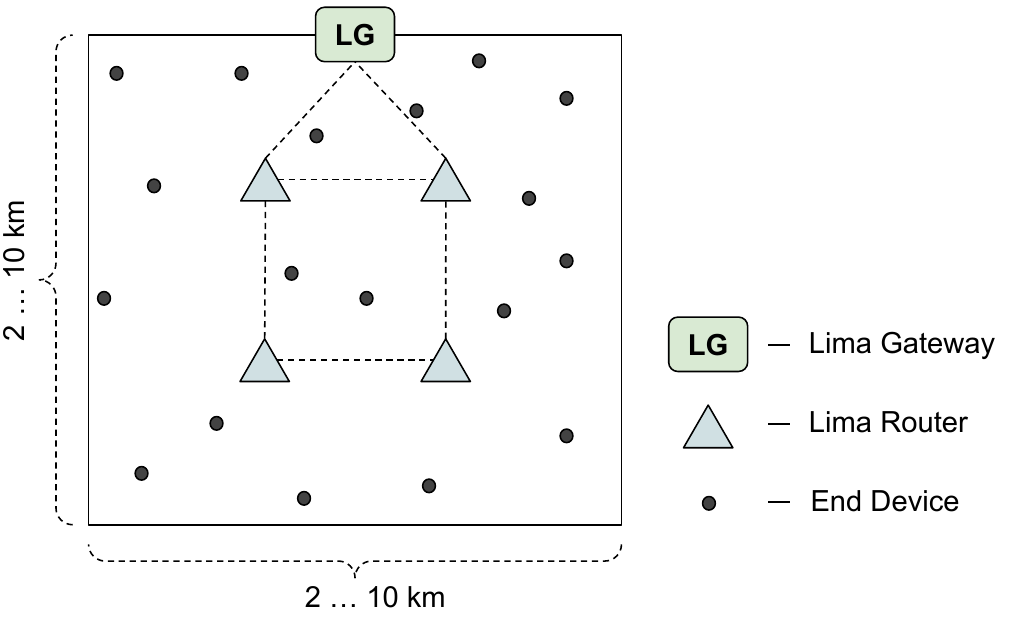}
\caption{LIMA simulation topology with increasing square size from 2 to 10 km, and increasing number of LRs.}
% \vspace{-0.4cm}
\label{fig:lima-ns3-topology}
\end{figure}

The simulation topology, depicted in Figure~\ref{fig:lima-ns3-topology}, consists of static End Devices (ED), LIMA Routers (LR) and a LIMA Gateway (LG), placed in a square area. A single LG is placed in the middle of the top edge of the square, with a varying number of LRs placed in a Manhattan-grid allocation on the square. End-devices are placed randomly within the square. The goal is to roughly model a precision agriculture or remote monitoring scenario with a number of sensors, a single gateway at the edge of the field and LIMA routers providing (enhanced) connectivity.

The number of end-devices is calculated based on a node density of \textit{1 ED per sq.~km}, so that there is a \textit{minimal sufficient amount} of LRs to cover the area while forming a connected LR network.

%%%+++%%%
Given that LRs are always placed in NxN grid, Table~\ref{tab:lima-lr-number} shows the least amount of LRs needed for a particular number of EDs, considering a density of 1 \textit{ED per sq. km}.

\begin{table}[ht]
\centering
\caption{Number of LRs used for given number of EDs, placed with a density of 1 \textit{ED per sq. km}.}
\vspace{-0.2cm}
\begin{tabular}{|c|c|}
\hline
\textbf{Number of EDs} & \textbf{Number of LRs used} \\ \hline
4                      & 1                           \\ \hline
6 - 16                 & 4                           \\ \hline
20 - 36                & 9                           \\ \hline
42 - 64                & 16                          \\ \hline
72 - 100               & 25                          \\ \hline
\end{tabular}
\label{tab:lima-lr-number}
\end{table}
%%%+++%%%

The evaluation was conducted over two scenarios. In the first scenario, the application traffic was fixed to 1 message every 1800 seconds from each ED to the gateway, with variable area sides that gradually increased from 2 to 10 \textit{km} in steps of 0.5 \textit{km} (see Figure~\ref{fig:lima-ns3-topology}). In the second scenario, the application traffic was increased while keeping a constant 6x6 \textit{km} area. Thus, we call the first scenario as \textit{Variable Size} (VS), and the second scenario as \textit{Variable Traffic} (VT). A summary of simulation parameters can be found in Table~\ref{tab:lima-sim-parameters}.

\begin{table}[t]
\centering
\vspace{-0.2cm}
\caption{Simulation parameters.}
\vspace{-0.3cm}
\label{tab:lima-sim-parameters}

\scriptsize
\setlength{\tabcolsep}{3pt}
\renewcommand{\arraystretch}{1.15}

\begin{tabularx}{\columnwidth}{@{}>{\centering\arraybackslash}p{0.47\columnwidth}>{\centering\arraybackslash}X@{}}
\toprule
\textbf{Simulation time, hours}             & 200                                                     \\
\textbf{Traffic period, seconds}            & 1800 for VS, from 7200 to 300 for VT                     \\
\textbf{Adaptive Data Rate (ADR)}           & Enabled                                                 \\
\textbf{Do Not Forward (DNoF) feature}      & Enabled                                                 \\
\textbf{Designated Router feature}          & Enabled                                                 \\
\textbf{Packet size, bytes}                 & 40                                                      \\
\textbf{Density, nodes per sq.\ km.}        & 1                                                       \\
\textbf{Area side, km}                      & 2-10 km with 0.5 steps, fixed to 6 km for VT  \\
\textbf{Min.\ distance between LRs, meters} & 1500                                                    \\
\textbf{Number of LGs}                      & 1                                                       \\
\textbf{Number of LRs}                      & 1--25 for VS, 9 for VT                                   \\
\textbf{Number of EDs}                      & 4--100 for VS, 36 for VT                                  \\
\bottomrule
\end{tabularx}

\vspace{-0.4cm}
\end{table}

For each scenario, four metrics were studied:

{\setlength{\leftmargini}{1.2em}
\begin{itemize}
\item \textit{Packet Delivery Ratio (PDR), \%}: The percentage of delivered packets relative to the number of sent packets.
\item \textit{Energy per ED, Joules}: The average amount of energy consumed by an ED over the duration of the simulation.
\item \textit{Packet Latency, ms}: The average E2E latency of a packet sent from an ED and delivered to the Network Server.
\item \textit{Energy per LR, Joules}: The average amount of energy consumed by an LR while forwarding traffic, including control messages.
\end{itemize}
}

\subsubsection{Simulation Results}

We present two sets of results, first for variable size (number of EDs) and fixed traffic load, and the other for variable traffic load and fixed size. We assume that LRs are placed at a fixed range from each other, in particular at the \textit{STP} range (currently SF7). This implies that the number of LRs will increase with the size and the area.

% var-size results
\begin{figure*}
\centering
\begin{minipage}{0.49\linewidth}
  \centering
\includegraphics[width=0.9\linewidth]{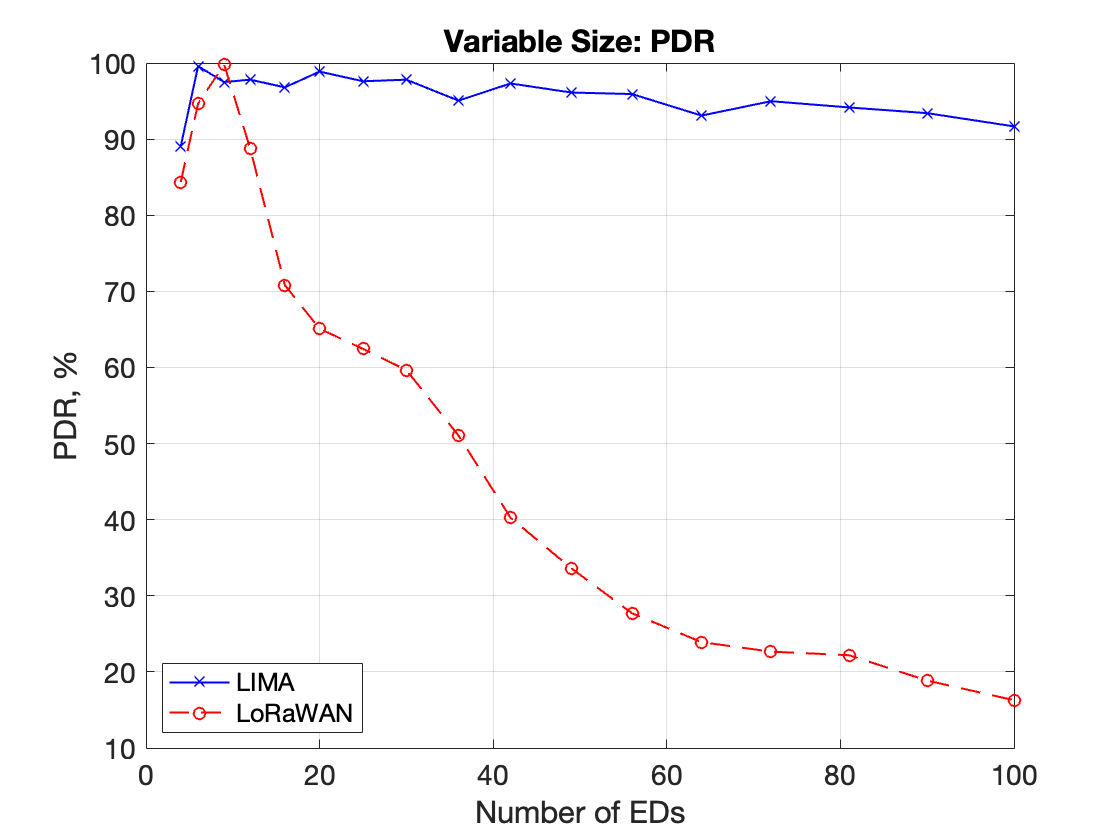}
  \caption{Variable Size: PDR, \%}
  \label{fig:var_size_pdr}
\end{minipage}\;\;%
\begin{minipage}{0.49\linewidth}
  \centering
\includegraphics[width=0.9\linewidth]{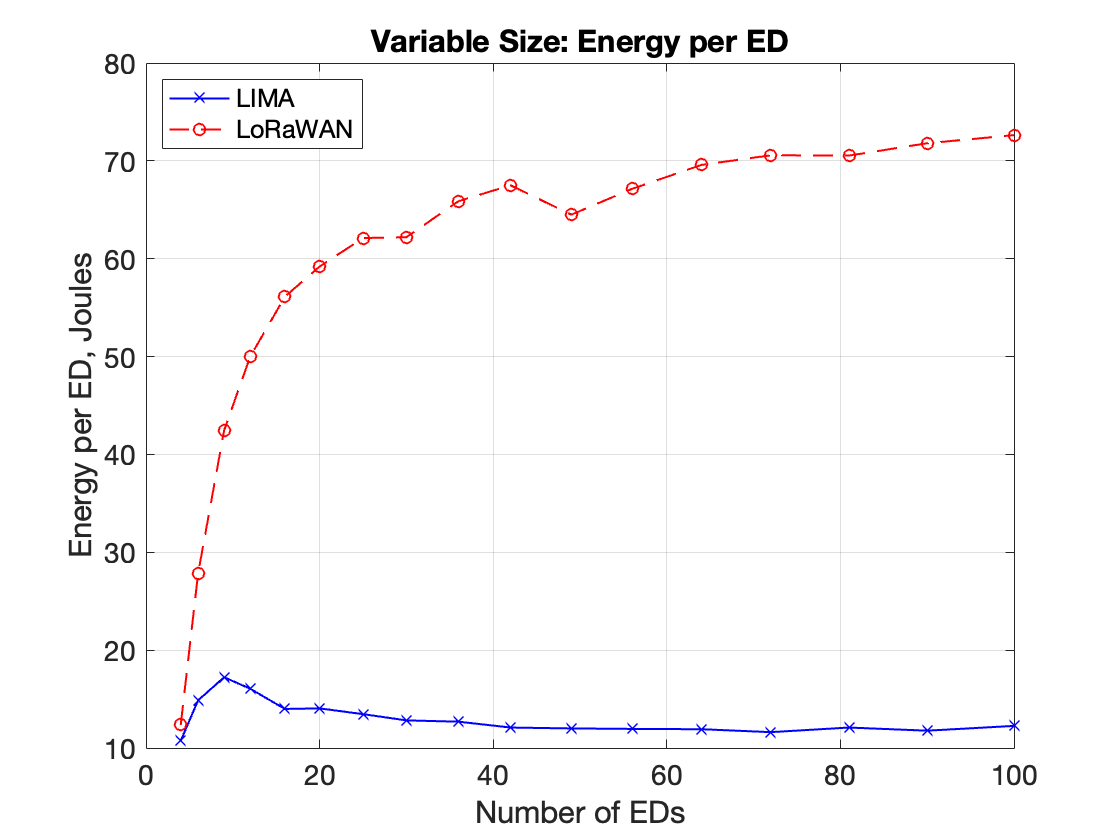}
  \caption{Variable Size: Energy per ED, Joules}
  \label{fig:var_size_energy}
\end{minipage}
\bigskip
\begin{minipage}{0.49\linewidth}
  \centering
\includegraphics[width=0.9\linewidth]{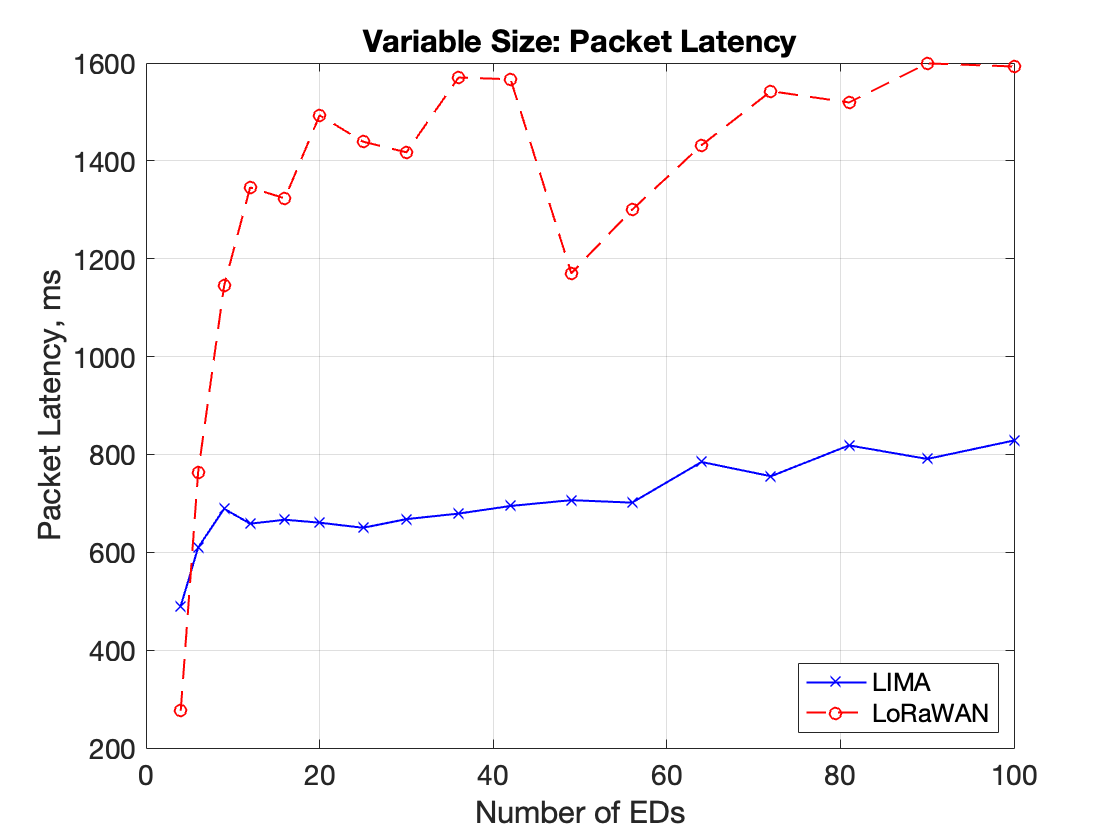}
\caption{Variable Size: Packet Latency, ms}  \label{fig:var_size_latency}
\end{minipage}\;\;
\begin{minipage}{0.49\linewidth}
  \centering
\includegraphics[width=0.9\linewidth]{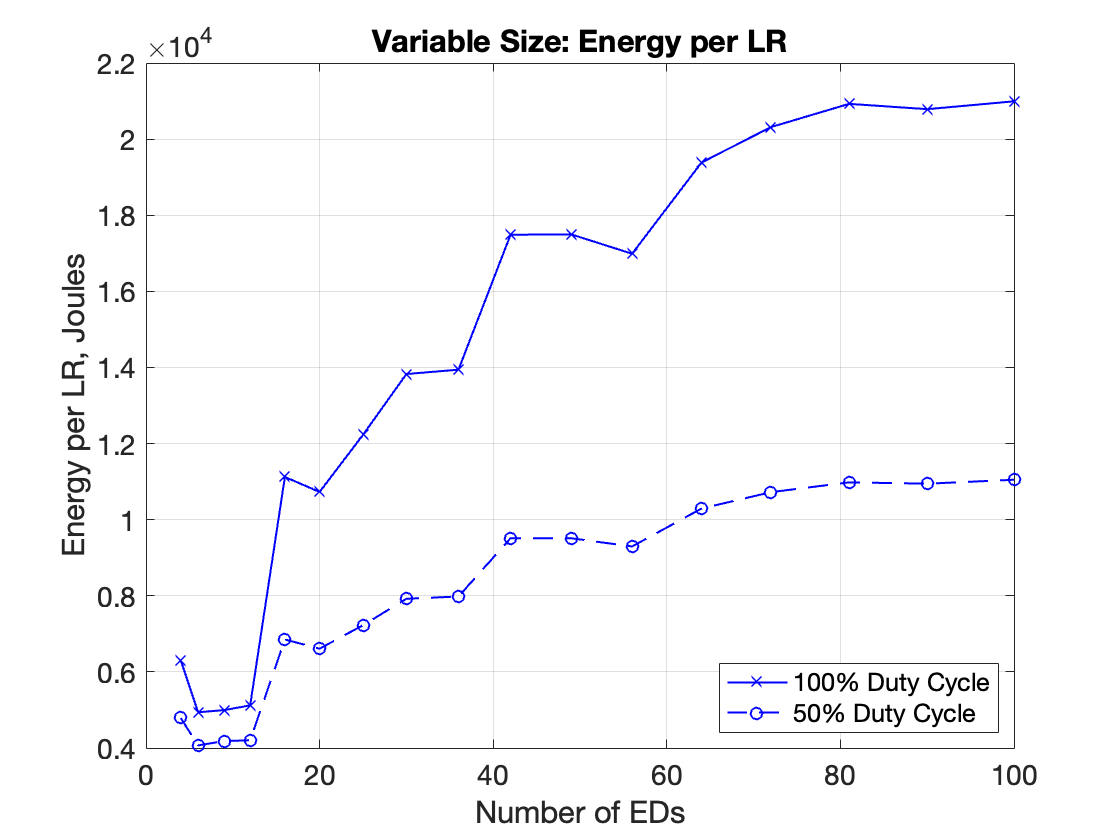}
\caption{Variable Size: Energy per LR, Joules}
\label{fig:var_size_energy_lr}
\end{minipage}
\vspace{-0.4cm}
\end{figure*}

% % new 1-row format
% \begin{figure*}[t]
% \centering
% % \captionsetup{font=small,skip=2pt}
% \captionsetup[sub]{font=footnotesize,skip=1pt}

% \begin{subfigure}[t]{0.25\textwidth}
%   \centering
%   \includegraphics[width=\linewidth]{Figures/var_size_pdr.png}
%   \subcaption{PDR, \%}
%   \label{fig:var_size_pdr}
% \end{subfigure}\hfill
% \begin{subfigure}[t]{0.25\textwidth}
%   \centering
%   \includegraphics[width=\linewidth]{Figures/var_size_energy.png}
%   \subcaption{ED energy, J}
%   \label{fig:var_size_energy}
% \end{subfigure}\hfill
% \begin{subfigure}[t]{0.25\textwidth}
%   \centering
%   \includegraphics[width=\linewidth]{Figures/var_size_latency.png}
%   \subcaption{Latency, ms}
%   \label{fig:var_size_latency}
% \end{subfigure}\hfill
% \begin{subfigure}[t]{0.25\textwidth}
%   \centering
%   \includegraphics[width=\linewidth]{Figures/var_size_energy_lr.png}
%   \subcaption{LR energy, J}
%   \label{fig:var_size_energy_lr}
% \end{subfigure}

% \vspace{-0.2cm}
% \caption{Variable Size results.}
% \label{fig:var_size_results}
% \vspace{-3mm}
% \end{figure*}

%var-traffic results
\begin{figure*}
\centering
\begin{minipage}{0.49\linewidth}
  \centering
\includegraphics[width=0.9\linewidth]{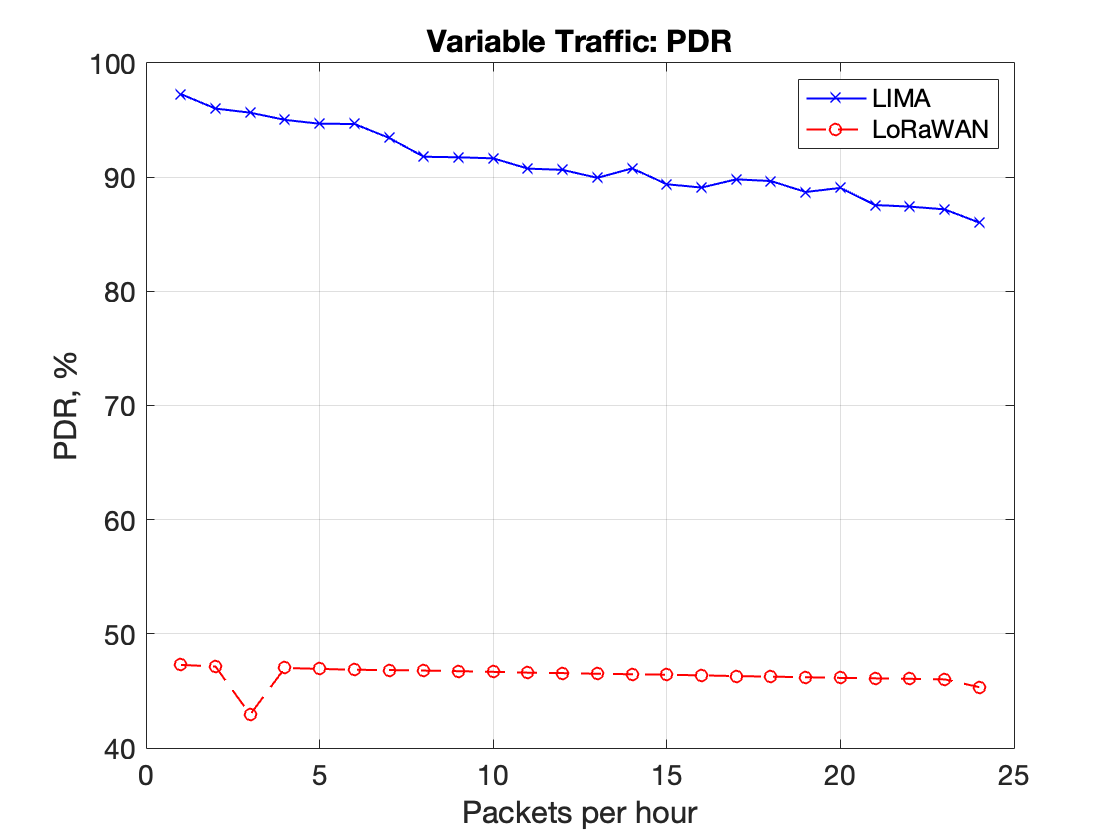}
  \caption{Variable Traffic: PDR, \%}
  \label{fig:var_traffic_pdr}
\end{minipage}\;\;%
\begin{minipage}{0.49\linewidth}
  \centering
\includegraphics[width=0.9\linewidth]{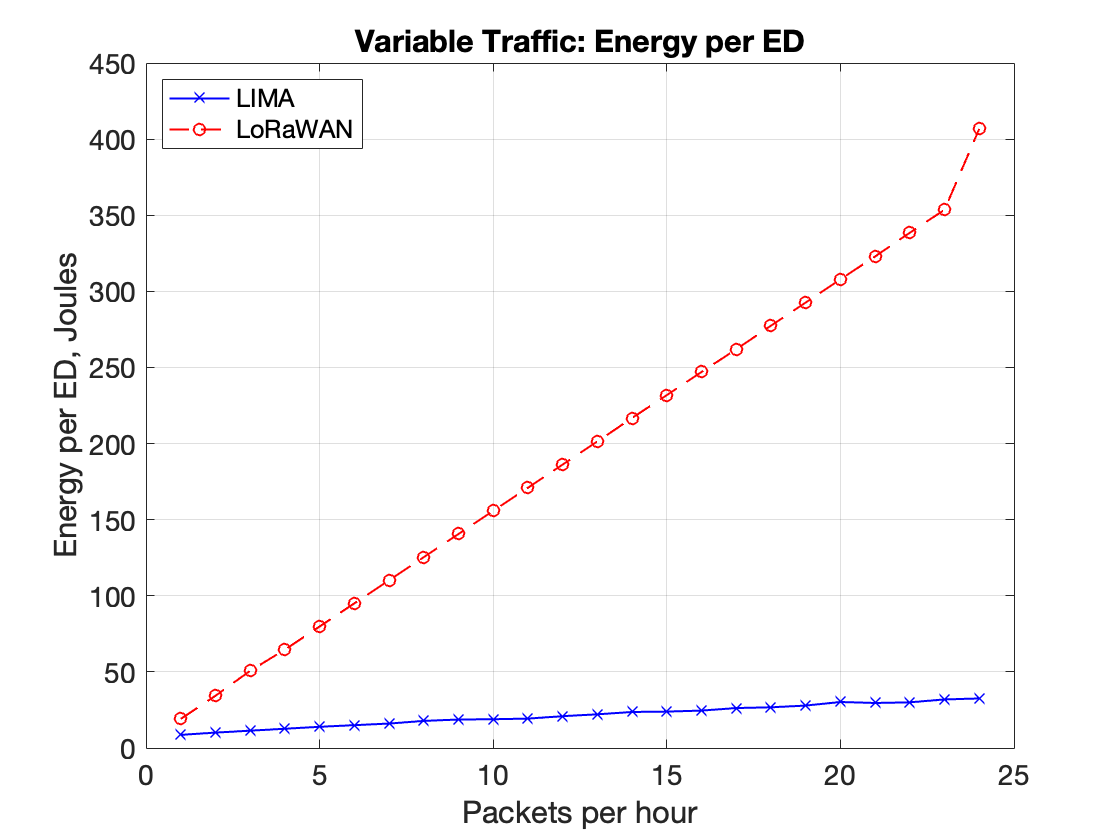}
  \caption{Variable Traffic: Energy per ED, Joules}
  \label{fig:var_traffic_energy}
\end{minipage}
\bigskip
\begin{minipage}{0.49\linewidth}
  \centering
\includegraphics[width=0.9\linewidth]{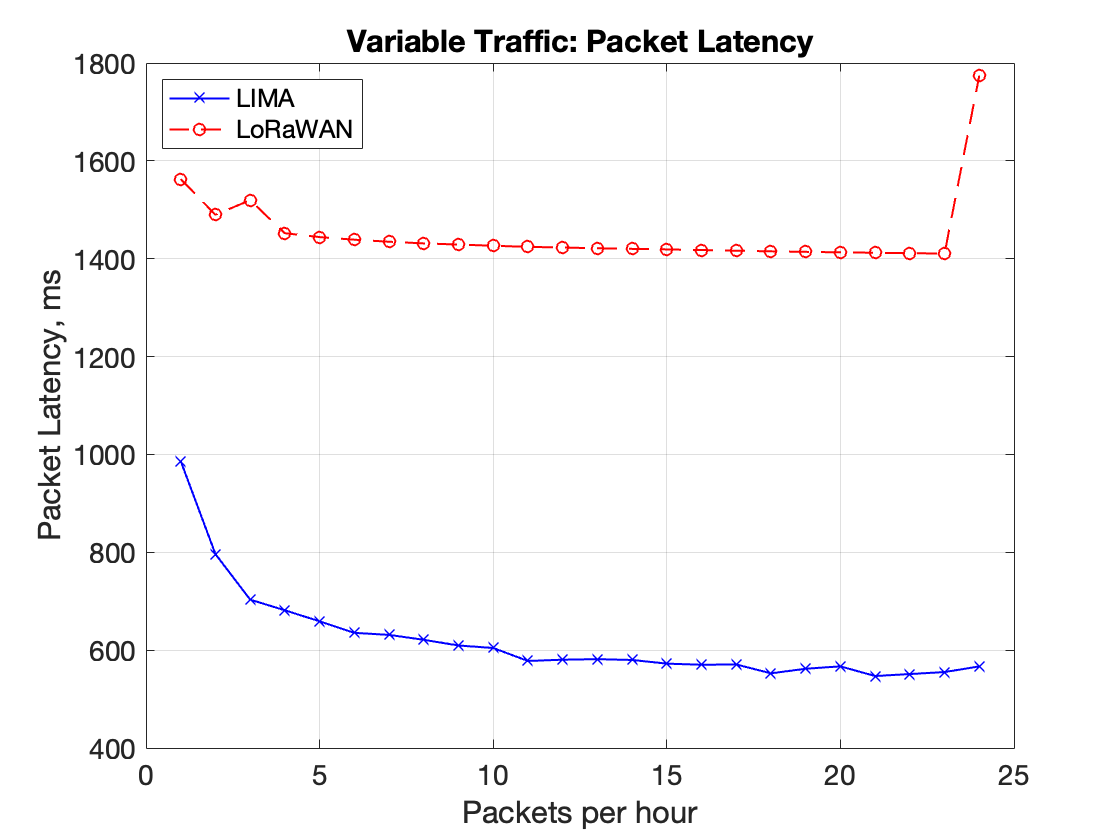}
\caption{Variable Traffic: Packet Latency, ms}  \label{fig:var_traffic_latency}
\end{minipage}\;\;
\begin{minipage}{0.49\linewidth}
  \centering
\includegraphics[width=0.9\linewidth]{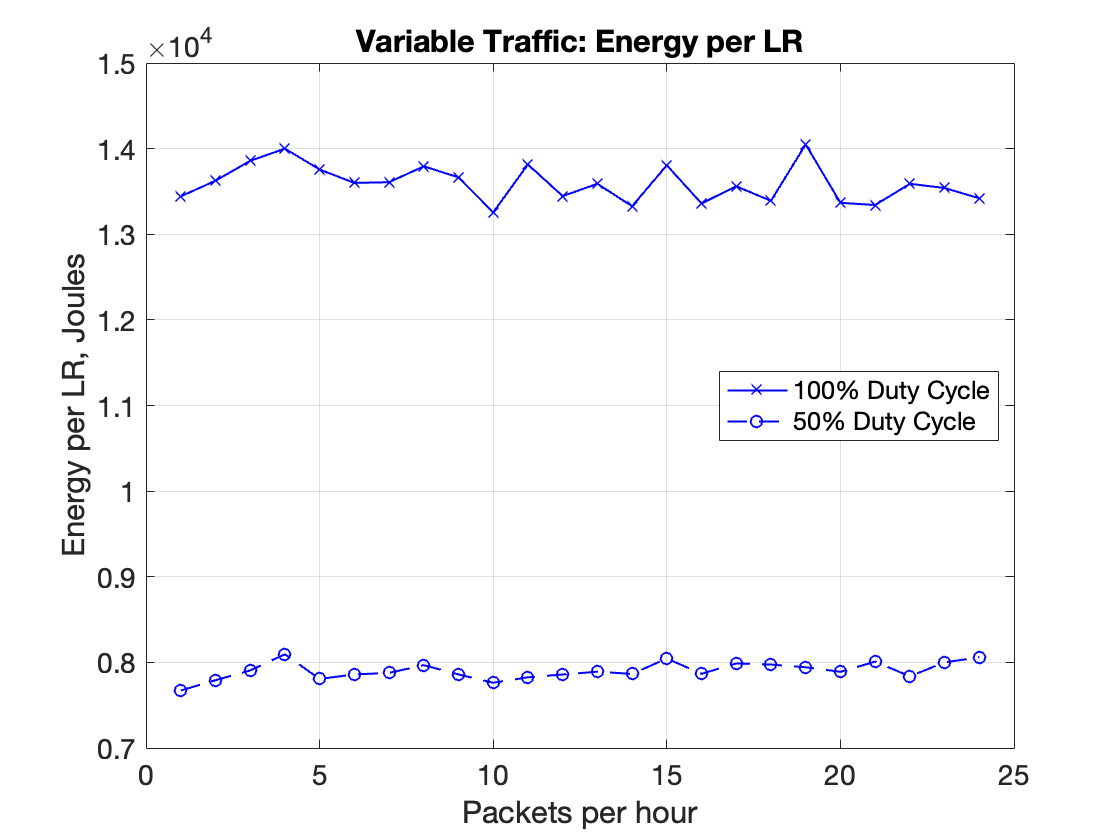}
\caption{Variable Traffic: Energy per LR, Joules}
\label{fig:var_traffic_energy_lr}
\end{minipage}
\vspace{-0.4cm}
\end{figure*}

% % new 1-row format
% \begin{figure*}[t]
% \centering
% % \captionsetup{font=small,skip=2pt}
% \captionsetup[sub]{font=footnotesize,skip=1pt}

% \begin{subfigure}[t]{0.25\textwidth}
%   \centering
%   \includegraphics[width=\linewidth]{Figures/var_traffic_pdr.png}
%   \subcaption{PDR, \%}
%   \label{fig:var_traffic_pdr}
% \end{subfigure}\hfill
% \begin{subfigure}[t]{0.25\textwidth}
%   \centering
%   \includegraphics[width=\linewidth]{Figures/var_traffic_energy.png}
%   \subcaption{ED energy (J)}
%   \label{fig:var_traffic_energy}
% \end{subfigure}\hfill
% \begin{subfigure}[t]{0.25\textwidth}
%   \centering
%   \includegraphics[width=\linewidth]{Figures/var_traffic_latency.png}
%   \subcaption{Latency (ms)}
%   \label{fig:var_traffic_latency}
% \end{subfigure}\hfill
% \begin{subfigure}[t]{0.25\textwidth}
%   \centering
%   \includegraphics[width=\linewidth]{Figures/var_traffic_energy_lr.png}
%   \subcaption{LR energy (J)}
%   \label{fig:var_traffic_energy_lr}
% \end{subfigure}

% \vspace{-0.2cm}
% \caption{Variable Traffic results.}
% \label{fig:var_traffic_results}
% % \vspace{-2mm}
% \end{figure*}

Figures~\ref{fig:var_size_pdr}~-~\ref{fig:var_size_energy_lr} show results for the Variable Size scenario with \textit{fixed density} of 1 ED/sq km. It is important to note for these figures that due to the constant density assumption, the X-axis represents not only increasing EDs, but also increasing area.
As shown on Figure~\ref{fig:var_size_pdr}, the PDR for the standard LoRaWAN case (with no LRs) rapidly decreases as the area grows. This is because as the number of EDs increases, so does the area, and an increasing number of EDs either have to communicate at higher SFs or to get out of range of the gateway. This is explained by a maximum transmission range limitation of LoRaWAN, that reaches a limit at SF12 and is approximately of 3 kilometers in the simulation model. With LIMA, the LRs effectively increase the coverage area, allowing EDs to both multi-hop to the LG as well as use SF7, thereby increasing Packet Delivery Ratio.

Figure~\ref{fig:var_size_energy} shows the average energy consumed by EDs. As the network size increases, the direct distance from ED to the gateway also increases; thus, it takes gradually more energy for EDs to deliver packets directly, as EDs use higher Spreading Factor and power to maintain connectivity. With LIMA, intermediate LRs help forward packets from EDs to gateway in a multi-hop fashion and the tunneled Adaptive Data Rate results in an ED using the SF and power to reach its closest LR. Since the distance from ED to the closest LR is much shorter than to the gateway, it takes significantly less energy for the ED to send out a packet.

From the message latency results shown in Figure~\ref{fig:var_size_latency}, it can be seen that LIMA is also more beneficial than the standard LoRaWAN in terms of latency. This, again, can be explained by the distance differences between direct ED-gateway vs ED-LR communication. Since the bigger distance results in a larger spreading factor (SF) employed by EDs to reach the gateway, the overall packet “on-air” duration significantly increases. For instance, with the short-range SF7 the air-time of a 40-byte packet is around 100ms, while the same packet size takes around 2500ms to be transmitted at SF12. Thus, even though the messages may go multiple hops, each hop takes 25x less time and therefore the overall latency is reduced.

Figure~\ref{fig:var_size_energy_lr} shows the average amount of energy consumed over the entire simulation duration of 200 hours by an LR while forwarding packets from EDs, both for the default 100\% duty cycle and for 50\% duty cycle. 
In both cases, the average energy per LR increases as expected, since more packets are being forwarded and received. 
The “jagged” nature of the curve with spikes is likely due to the fact that the number of LRs increase in a step-wise manner as the number of EDs is increased. 

Figures~\ref{fig:var_traffic_pdr}~-~\ref{fig:var_traffic_energy_lr} show results for the Variable Traffic scenario. 
Figure~\ref{fig:var_traffic_pdr} shows the PDR as traffic increases. Again, LIMA tends to maintain significantly higher PDR compared to the standard LoRaWAN case since LIMA relies on the packet forwarding via LRs, while pure LoRaWAN attempts to reach the direct uplink communication with the gateway.

As for the energy per ED metric in the Variable Traffic scenario, it grows linearly with the increased traffic rate, as shown on Figure~\ref{fig:var_traffic_energy}. LIMA manages to save a substantial amount of ED's energy by employing intermediate LRs to shorten an effective communication distance for the EDs. In the standard LoRaWAN case, significantly more energy is required to reach the gateway directly.

Figure~\ref{fig:var_traffic_latency} shows the average latency with the increased traffic rate. It can be noticed that for both LIMA and LoRaWAN cases, the average packet latency is changing insignificantly, since the network topology does not change in terms of number of EDs and LRs. However, with the increasing traffic, more packets get lost either due to collisions or duty-cycle limitations, which contributes to the spike in the LoRaWAN curve at around 23 packets per hour.
%but this does not affect the overall latency, since the unconfirmed packets are being transmitted.

The average energy consumed by an LR with the increasing traffic rate also grows, as shown on Figure~\ref{fig:var_traffic_energy_lr}. For both duty cycles, the trend is more or less flat. This is likely due to the fact that the load at the traffic rate and ED size is not challenging, and much of this traffic may not have to go through the LR in the 6x6 km deployment used. Consequently, the energy consumption is dominated by the idle state, which is not affected by traffic and therefore results in a flat trend.

Our findings are summarized below. For constant traffic (Variable Size), LIMA outperforms LoRaWAN by several factors in terms of PDR, scalability, energy savings and latency. Specifically:

{\setlength{\leftmargini}{1.2em}
\begin{itemize}
    \item At 100 EDs, LIMA PDR is >5x higher than LoRaWAN PDR.
    \item The number of devices supportable at a 85\% PDR threshold is >8x more than LoRaWAN.
    \item LIMA ED energy consumption at 20 and 100 EDs is >4x and >6x less than LoRaWAN respectively.
    \item LIMA latency at 20 and 100 EDs is >2.3x and >2x less than LoRaWAN respectively.
\end{itemize}
}

For constant ED number (Variable Traffic), LIMA continues to outperform LoRaWAN -- more modestly on PDR, but more impressively on Energy. Specifically:

{\setlength{\leftmargini}{1.2em}
\begin{itemize}
    \item As traffic increases from 1 pkt/hour to 24 pkts/hour, LIMA PDR outperformance is relatively the same -- from >2x to >1.86x.
    \item As traffic increases from 1 pkt/hour to 24 pkts/hour, the LIMA energy consumption goes from nearly the same to as much as 12.6x lower.
    \item As traffic increases from 1 pkt/hour to 24 pkts/hour, the LIMA packet latency goes from 1.63x lower to 3.5x lower.
\end{itemize}
}

The average transmit energy expended by a LIMA Router (LR) is very modest, less than 22000 Joules over the entire simulation time of 720,000 seconds (30 mW) in any scenario. Thus, even accounting for other sources of energy drain such as CPU and memory, an LR can be easily supported by inexpensive and light 10W solar panels, assuming enough sunlight. Further, we observe that the energy per LR only increases gradually with size, and is quite stable with traffic. Specifically:

{\setlength{\leftmargini}{1.2em}
\begin{itemize}
    \item As the number of EDs increases from 4 to 100, the energy consumption per LR decreases modestly by a factor of about 2.5x.
    \item As the traffic increases from 1 to 24 pkts/hour, the energy consumption increases modestly by a factor of about 1.18x.
\end{itemize}
}

%%%+++%%%
Based on these numbers, we conclude that LIMA has the potential to provide a dramatic boost to the performance of LoRaWAN in many realistic scenarios.
%%%+++%%%

\subsection{LIMA Prototype and Experiments}
\label{sec:prototype}
In this section, we describe the construction of a LIMA prototype using a commercial LoRaWAN gateway as the starting point, and describe experiments to validate its operation. While the purpose of the simulation described in the previous section was to validate the design and evaluate performance at scale, the goal with the prototype and real-world testing is to confirm that the LIMA data plane tunneling and frames structure is compatible with LoRaWAN, that authentication works, and that multi-hop range extension and energy savings can be achieved without modifying the end-device or Server, and quantify any such gains. 

Accordingly, the prototype network is minimal, with one LG, one LR and one ED, which tests a maximum of two hops. The key operational risks in more hops show up mostly in the \textit{control plane} (routing etc.) where the interaction with LoRaWAN specifics is minimal, and so has been validated by high-fidelity simulations. 
Further, the processes of packet encapsulation, inspection, and forwarding are functionally identical at each hop (Figures 4, 5), the 1-hop prototype can serve as a proof-of-concept for fundamental compatibility with LoRaWAN. Finally, since the goal of the prototype is to evaluate the data plane and not the control plane, it uses static routes instead of the full-fledged REM based routing.

\subsubsection{Prototype Components}
\label{sec:prototype-components}
The prototype assembly components include:

{\setlength{\leftmargini}{1.2em}
\begin{itemize}
\item Dragino LSN50v2 LoRaWAN temperature and humidity sensor.
\item Seeed Studio WM1302 LoRaWAN Gateway module, based on the Semtech SX1302 baseband LoRa chip. 
\item Seeed Studio Raspberry Pi HAT board for connecting the Seeed Studio WM1302 LoRaWAN Gateway module to Raspberry Pi.
\item Raspberry Pi single board computers.
\item Connectors, isolation boxes and other miscellaneous.
\item The Things Network (TTN) LoRaWAN Server.
\end{itemize}
}

The Dragino sensor is used as the End Device (ED), and the Seeed Studio LoRaWAN Gateway module is used for both the LIMA Gateway (LG) and the LIMA Router (LR). Modification of existing code provided by Seeed Studio running on a Raspberry Pi attached to the Gateway module implements the LIMA Router or LIMA Gateway functionality as the case may be. 

The LR functionality is achieved by starting with the vanilla code for a LoRaWAN Gateway, disabling network-server specific functions, and including the LIMA protocol described in the preceding sections. Tunneled ADR, as well as receive window management is implemented. As mentioned earlier, in accordance with the focus of the prototype evaluation as a complement to simulations, static routes are used, obviating the need for designated edge routers. The DNoF functions are also dispensed with given the small size of the network and in line with the goals of the prototype evaluation.

The LG functionality is very close to the original LoRaWAN Gateway functionality, except that it strips off the LIMA header from uplink LIMA packets and recovers the original LoRaWAN packets before sending the packet to the network server. Similarly, it prepends the LIMA header to the downlink LoRaWAN packets. It also updates the metadata sent to the server using the ADR information in the LIMA header as described in section \ref{sec:tunneled-adaptive-data-rate}.

No modifications were made to the Dragino sensor end device, or the TTN LoRaWAN server.

\subsubsection{Indoor Attenuator-based Experiments}

\begin{figure}[h]
\centering
\vspace{-0.2cm}
\includegraphics[width=0.6\linewidth]{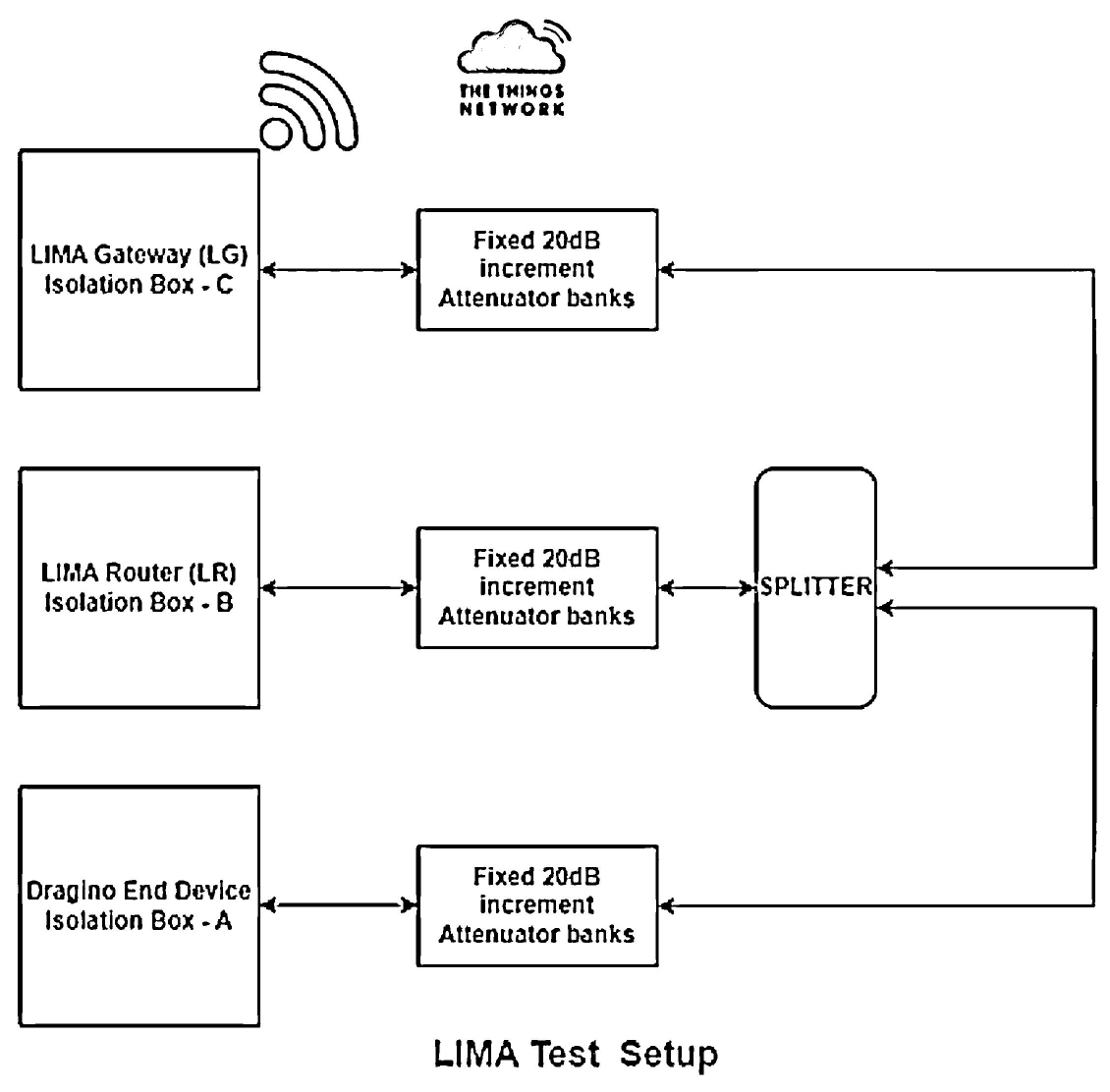}
\vspace{-0.2cm}
\caption{Indoor test configuration for LIMA testing. For LoRaWAN the ED connects to the LG directly.}
\label{fig:lima-test-setup}
\vspace{-0.3cm}
\end{figure}

We have compared the performance between LoRaWAN and LIMA in a conducted Faraday cage lab test setup.
%%%+++%%%
We use three Ramsey STE2200 isolation boxes (faraday cage), a power splitter, and fixed 20 dB attenuators that can be combined to provide the desired attenuation between the ED/LR/LG.
%%%+++%%%
The testing set up for LIMA is illustrated in Figure~\ref{fig:lima-test-setup}. The LoRaWAN setup simply excludes the middle LR and connects the ED to the LG directly via the splitter. 

Using this setup, we experimented with various combinations of attenuations between the ED, LR and LG for LIMA, and between the ED and LG for vanilla LoRaWAN. For each combination, 100 messages were sent from the Dragino sensor and we measured a number of data points as shown in Table~\ref{tab:attenuation-results}.

\begin{table*}[t]
\centering
\vspace{-0.2cm}
\caption{Indoor test configuration results (LIMA vs. LoRaWAN).}
\vspace{-0.2cm}
\label{tab:attenuation-results}

\scriptsize
\setlength{\tabcolsep}{3pt}
\renewcommand{\arraystretch}{1.15}

\resizebox{\textwidth}{!}{%
\begin{tabular}{cccccccccccccc}
\toprule
\multicolumn{14}{c}{\textbf{LIMA Test Results}} \\
\midrule
\multicolumn{2}{c}{\textbf{Attenuation (dB)}} &
\multirow{2}{*}{\textbf{Total Atten. (dB)}} &
\multicolumn{3}{c}{\textbf{DR - TTN (UL)}} &
\multicolumn{2}{c}{\textbf{ADR Message (DL)}} &
\multicolumn{3}{c}{\textbf{RSSI (dB)}} &
\multirow{2}{*}{\textbf{Tx pkts by ED}} &
\multirow{2}{*}{\textbf{Rx by TTN}} &
\multirow{2}{*}{\textbf{TTN PDR \%}} \\
\cmidrule(lr){1-2}\cmidrule(lr){4-6}\cmidrule(lr){7-8}\cmidrule(lr){9-11}
\textbf{ED to LR} & \textbf{LR to LG} & &
\textbf{BW (kHz)} & \textbf{SF} & \textbf{CR} &
\textbf{TX DR} & \textbf{Tx Pwr (dBm)} &
\textbf{@ ED} & \textbf{@ LR} & \textbf{@ LG} & & & \\
\midrule
80  & 20 & 100 & 125 & 7  & 4/5 & 3 & 2  & -83  & -98  & -19 & 76  & 68  & 89.47 \\
\textbf{100} & \textbf{20} & \textbf{120} & \textbf{125} & \textbf{7}  & \textbf{4/5} & \textbf{3} & \textbf{18} & \textbf{-102} & \textbf{-103} & \textbf{-16} & \textbf{115} & \textbf{114} & \textbf{99.13} \\
100 & 40 & 140 & 125 & 7  & 4/5 & 3 & 18 & -102 & -103 & -38 & 100 & 89  & 89.00 \\
100 & 60 & 160 & 125 & 7  & 4/5 & 3 & 18 & -102 & -104 & -57 & 100 & 89  & 89.00 \\
100 & 80 & 180 & 125 & 7  & 4/5 & 3 & 18 & -102 & -103 & -78 & 100 & 89  & 89.00 \\
120 & 80 & 200 & 125 & 10 & 4/5 & 0 & 30 & -120 & -123 & -78 & 101 & 91  & 90.10 \\
140 & 80 & \textcolor{red}{\textbf{220}} &
\multicolumn{11}{c}{NO JOIN, NO PACKETS SEEN BY LR/LG, TOO MUCH ATTENUATION} \\
\midrule
\multicolumn{14}{c}{\textbf{LoRaWAN Test Results}} \\
\midrule
0   &  & 0   & 127 & 7  & 4/5 & 3 & 2  & -83  &  & -19 & 76  & 68  & 89.47 \\
80  &  & 80  & 125 & 7  & 4/5 & 3 & 2  & -102 &  & -16 & 115 & 114 & 99.13 \\
100 &  & 100 & 125 & 7  & 4/5 & 3 & 14 & -102 &  & -38 & 100 & 89  & 89.00 \\
\textbf{120} &  & \textbf{120} & \textbf{125} & \textbf{10} & \textbf{4/5} & \textbf{0} & \textbf{30} & \textbf{-102} &  & \textbf{-57} & \textbf{100} & \textbf{89}  & \textbf{89.00} \\
140 &  & \textcolor{red}{\textbf{140}} &
\multicolumn{11}{c}{NO JOIN, NO PACKETS SEEN BY LR/LG, TOO MUCH ATTENUATION} \\
\bottomrule
\end{tabular}%
}

\vspace{0.25em}

{\footnotesize
\setlength{\parskip}{0pt}
\setlength{\baselineskip}{0.9\baselineskip}
\begin{minipage}{\textwidth}
\justifying
\noindent
LIMA can accommodate a total attenuation of 200 dB between the ED and Gateway vs only 120 dB for LoRaWAN. Further, at the 120 dB attenuation, the ED with LoRaWAN uses DR-0 (SF 10, 30 dBm) whereas LIMA's tunneled ADR results in ED using DR-3 (SF 7, 18 dBm) at the same attenuation, which is considerable energy savings.
\end{minipage}\par
}

\end{table*}

%%% BELOW IS DEPRECATED PNG VERSION OF THE TABLE, KEEPING AROUND IN CASE NEEDED
% \begin{figure}[h]
% \centering
% \includegraphics[width=0.8\linewidth]{Figures/attenuation-results.png}
% \caption{Indoor test configuration results. LIMA can accommodate a total attenuation of 200 dB between the ED and Gateway vs only 120 dB for LoRaWAN. Further, at the 120 dB attenuation, the ED with LoRaWAN uses DR-0 (SF 10, 30 dBm) whereas LIMA's tunneled ADR results in ED using DR-3 (SF 7, 18 dBm) at the same attenuation, which is considerable energy savings. The corresponding relevant fields are highlighted in yellow.}
% %\vspace{-0.4cm}
% \label{fig:attenuation-results}
% \end{figure}

The DR-TTN column shows the bandwidth, spreading factor and coding rate employed by the ED in its uplink transmission, as observed at the TTN server. The downlink ADR message instructions on the Data Rate (DR) index and power are shown in the next box -- equivalently, these are the DR and power values used by the ED\footnote{LoRaWAN sends the data rate and power values using indices which are mapped into actual data rate and power depending on the region (see \cite{loraallianceRP002104LoRaWAN}). For convenience, we have made the conversion, which is 30 dBm - 2*index and shown the \textit{actual power}}. The RSSI and message delivery statistics are shown on the right.

% In the Attenuation box, the green items are those total combinations which resulted in successful connectivity whereas the red items are those that were unsuccessful.

We observe that with the ED transmitting at full power (30 dBm), LoRaWAN can only handle a total attenuation of 120 dB between ED and NS compared to 200 dB for LIMA. We also observe that when the ED to LR attenuation is 100 dB and the LR to LG is 20 dB, the ED transmits at DR3 (US SF7) and 18 dBm. In comparison, for an equivalent attenuation of 120 dB in LoRaWan, the ED transmits at DR0 (US SF10) and 30 dBm. SF10 consumes about 6x power as SF7, and 30 dBm consumes significantly higher power than 18 dBm. These lines are highlighted in bold in the Table~\ref{tab:attenuation-results}. In fact, with LIMA the above holds true until a total attenuation of 180 dBm is reached, implying a longer range as well. 
Thus LIMA reduces the total energy consumption significantly. Further, as indicated by the last column, the PDR of LIMA is 10\% better than that of LoRaWAN for the same total attenuation as well. 

\subsubsection{Outdoor Experiments}

We conducted outdoor testing in downtown Brooklyn, New York. We conducted tests for both LoRaWAN and LIMA using the prototypes described in section \ref{sec:prototype-components}. 
We note that the ED and TTN server had no modifications whatsoever. 

We performed two tests, a range test and an energy test. The goal of these tests was to demonstrate that the use of LIMA enables an ED to communicate from a longer distance and with lower energy (power and spreading factor), without any changes to the ED or the TTN server, and in a LoRa-rich interference environment.

\paragraph{Range Test.}

The range test was conducted as follows. The ED and LoRaWAN gateway were apriori registered with the TTN server and were started up on the 4th floor of a building. This is basically the LoRaWAN network, without any LIMA. 
%%%+++%%%
The TTN server logs were observed to show that the ED joined, and messages from the ED started appearing on the TTN console. The Gateway was positioned right next to the window.
%%%+++%%%

Two team members headed out, one with the ED and the other with the LR. The LR was turned off whereas the ED was turned on. The members moved away from the building until the connectivity was broken between the ED and LG, as observed from the TTN logs. We noted this limit of LoRaWAN range. 

%%%+++%%%
We observed that  as the ED was moved away, ADR messages ramped up the ED power, up to 28 dBm but the spreading factor (SF) was not changed, remained at 7. This may be because of lack of suitable downlink connectivity. We observed loss of several consecutive packets at the range limit.
%%%+++%%%

\begin{figure}[t]
\centering
\includegraphics[width=0.7\linewidth]{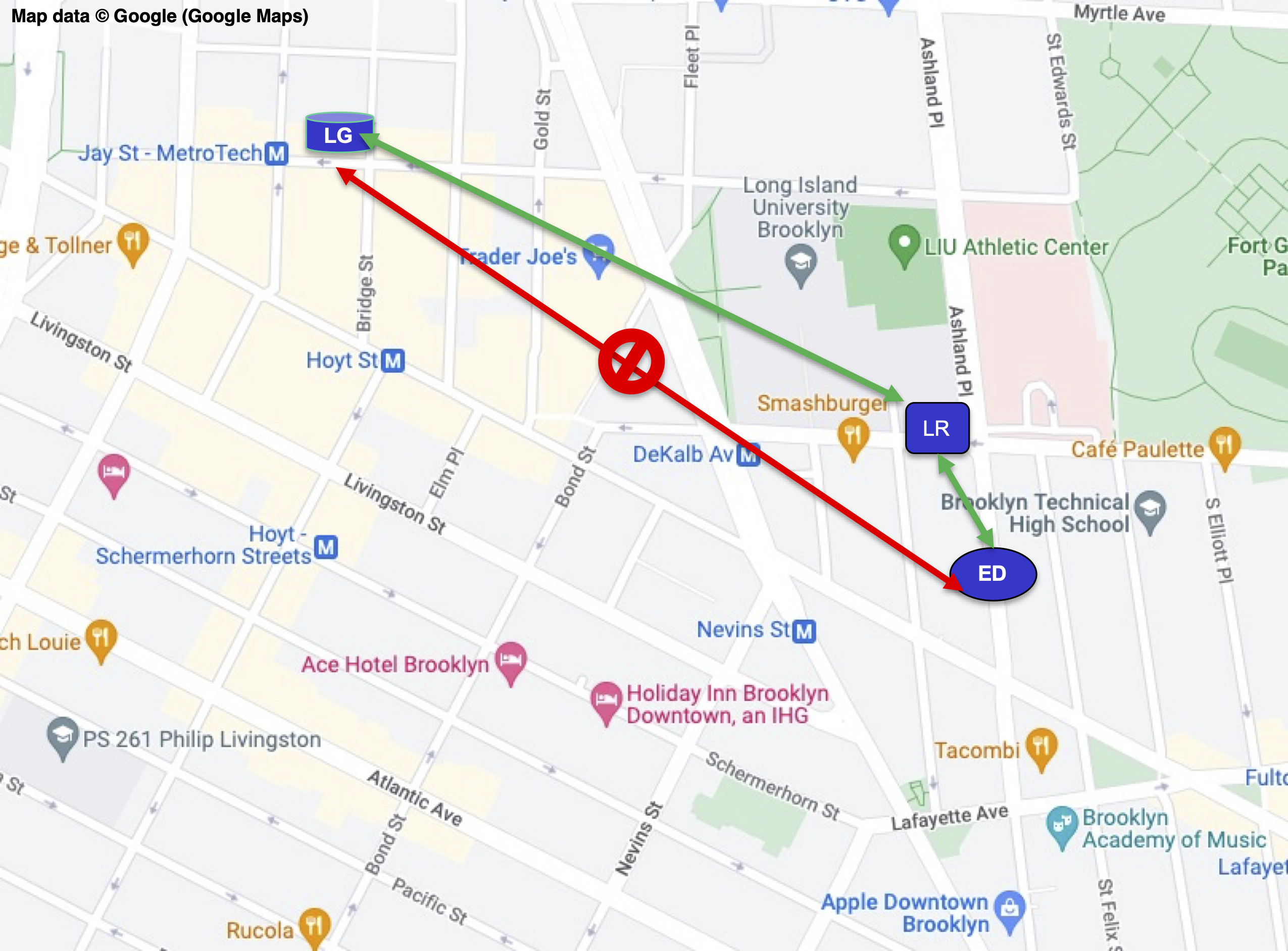}\par

{\footnotesize
\setlength{\parskip}{0pt}
\setlength{\baselineskip}{0.9\baselineskip}

\begin{minipage}{\linewidth}
\justifying
\noindent
The ED was unable to connect to the LG between the locations shown. When an LR was activated in the location shown, a multi-hop connection was confirmed.
\end{minipage}\par
}

\vspace{-0.6em}
\caption{Range Test.}
\label{fig:range-test}
\end{figure}

We then activated LIMA in the Gateway with a single command that activated LIMA code. That is, the LoRaWAN Gateway was transformed into a LIMA Gateway.
The team member carrying the LIMA Router (LR) moved to a point in between ED and LG while turning on the LR.
The connectivity was re-established, and the messages began to appear on the TTN logs, showing that the messages hopped via the LR. This continued to be the case when the ED was taken even further out. Logs on both the LR, LG showed uplink and downlink messages through the LR. The TTN messages continued stably for several minutes thereafter. Figure~\ref{fig:range-test} shows the locations of G/LG (Gateway in LoRaWAN mode and LG in LIMA mode), LR and ED when the ED was not able to connect to the server without the LR, but was able to do so via the LR when the LR was activated.

\paragraph{Energy Test.}

The energy test was conducted in a manner similar to the Range test, except that connectivity was stretched but not broken. Specifically, we first restarted all the components in LoRaWAN mode, and after the ED joined, two team members headed out, one with the ED and the other with the LR. The LR was turned off whereas the ED was turned on. The members moved away from the building. As the ED was moved away, the ADR messages ramped up its power. The ED stopped moving when the power was index 0 (transmit power 30 dBm), but connectivity was not broken.

Now LIMA was activated as before, that is, the LoRaWAN gateway was switched to operate as a LIMA Gateway with a single command that activated LIMA code. The team member carrying LR then moved to a point in between the ED and LG while turning on the LR.
We noticed that the Transmit Power index on the ED increased to 1, i.e, the transmit power dropped by 2 dBm from 30 dBm to 28 dBm. 
Attempts to further reduce the transmit power were not successful possibly due to higher ISM band noise in the vicinity of the ED which compromised downlink connectivity for ADRs, and due to the difficulty of positioning the LR appropriately. 

\begin{figure}[t]
\centering
\includegraphics[width=0.7\linewidth]{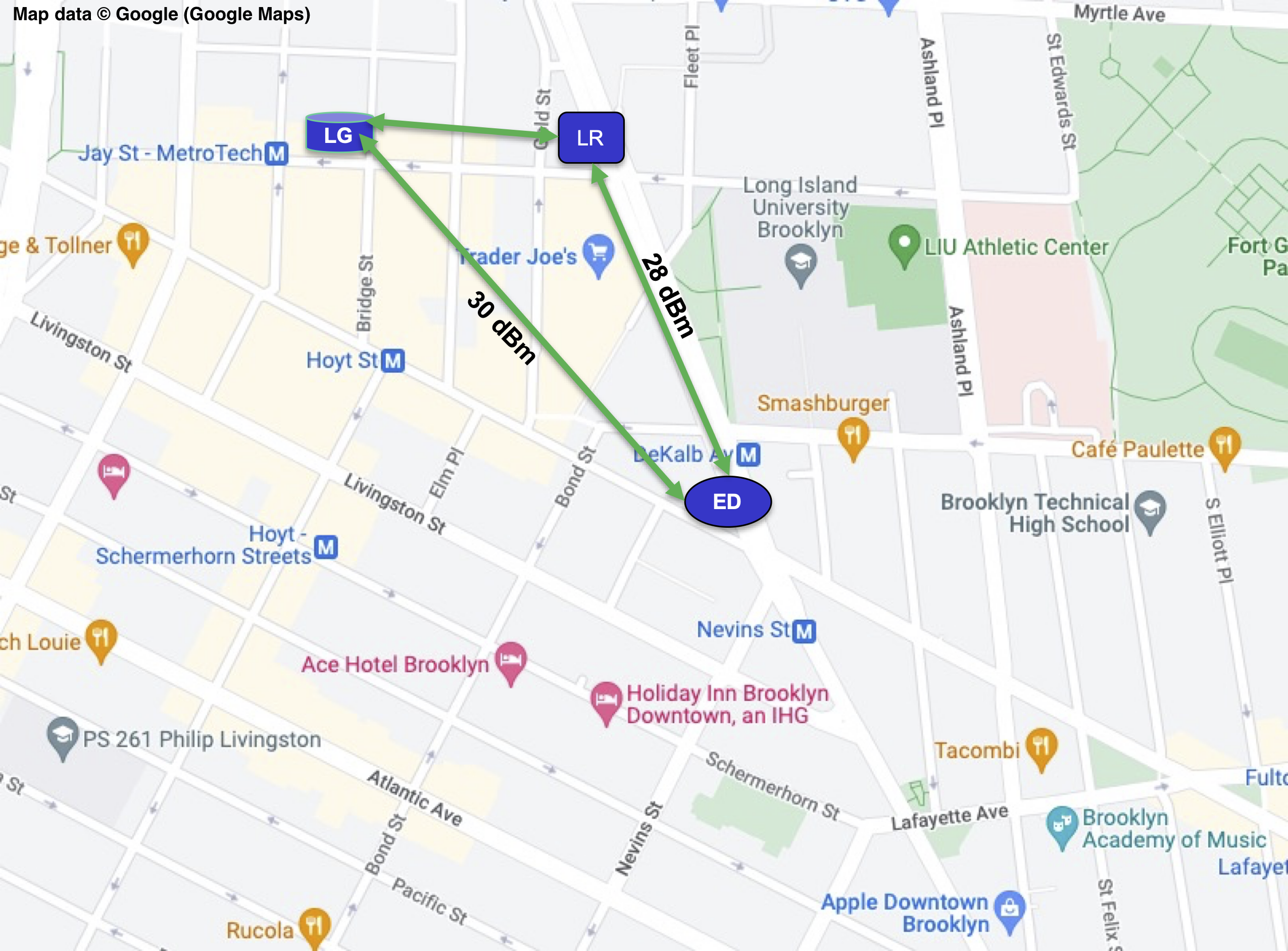}\par

{\footnotesize
\setlength{\parskip}{0pt}
\setlength{\baselineskip}{0.9\baselineskip}

\begin{minipage}{\linewidth}
\justifying
\noindent
With no LIMA Router (LR), the Adaptive Data Rate settled at 30 dBm power at the ED.
With the LR in place, the server reduced the ED's transmit power to 28 dBm.
\end{minipage}\par
}

\vspace{-0.6em}
\caption{Energy Test.}
\label{fig:energy-test}
\vspace{-0.4cm}
\end{figure}

Figure~\ref{fig:energy-test} shows the locations of the ED, LR and G/LG (Gateway in LoRaWAN mode and LIMA Gateway in LIMA mode) when a direct connection without the LR required 30 dBm of power, but only required 28 dBm via the LR.

Our prototype-based testing showed that:
{\setlength{\leftmargini}{1.2em}
\begin{itemize}
    \item Multi-hop relaying functionality can be introduced into a real-world LoRaWAN network without making any changes whatsoever to the end devices, the LoRaWAN server(s), or the LoRaWAN protocol.
    \item Multi-hop joining and authentication of end devices works in a transparent manner.
    \item The Adaptive Data Rate (ADR) can be extended to have the ED use only the power needed to reach its LIMA Router.
    \item LIMA with a single LR can extend the range of sensors by nearly 2x, reduce the transmit power by 12 dBm and packet duration by 6x (for equivalent range).
    %\item The LG-LR communication appears to be weaker than the LG-ED communication, ie, the latter tolerates higher attenuations (path loss) than the former.
\end{itemize}
}

%%%+++%%%
It should be noted that the tests were conducted in downtown Brooklyn -- an urban area, near a large university, with a lot of RF interference in the unlicensed bands. Indeed, we could detect numerous LoRaWAN end-devices and Gateways in the vicinity. LIMA was able to co-exist with other devices and gateways for a better part of a day in a LoRaWAN-rich environment. This means that the LIMA Routers should be able to be “dropped in” without causing any harm to existing networks. 

The above test demonstrated that LR was able to provide forwarding functionality. Extending this to a mesh network is not dependent upon LoRaWAN-specific issues, and hence is not considered risky. While the use of only one LR does not yield insights into performance at scale, the ns-3 simulation results discussed earlier (section~\ref{sec:simulation} show the results of a model with the full mesh networking.
%%%+++%%%

\section{Concluding Remarks}
\label{sec:concluding-remarks}
We have described LoRaWAN IoT Mesh Augmentation (LIMA), a protocol and system that provides robust multi-hop routing, range extension and energy savings for LoRaWAN end devices. Key features include highly efficient mesh routing based on path-reversal techniques; transparent Adaptive Data Rate (ADR) support based on the signal strength from an ED to the nearest LR; a Do-Not-Forward (DNoF) list to suppress duplicate transmissions; election of a designated LR for each ED; and multi-hop downlink ED window management. Compared to existing work, LIMA is unique in that it is the first scalable multi-hop mesh networking protocol that transparently augments LoRaWAN for uplink and downlink range-extension and end device energy reduction without requiring changes to end devices or the server(s).

Our evaluation of LIMA via simulation using ns-3 over a notional precision agriculture / remote monitoring scenario shows that it can provide multi-factor gains in delivery, scalability, latency and energy consumption. We have built a LIMA prototype by modifying a commercial LoRaWAN gateway product and evaluated it both on an attenuator-based indoor testbed and outdoors. Our prototype work confirms LIMA's real-world feasibility, and testing in a LoRa-rich environment shows expected behavior. We believe that LIMA can be a scalable, cost-efficient, and easy-to-deploy "drop in" solution for augmenting private as well as public LoRaWAN networks, and significantly extending the effective range and battery life of LoRaWAN devices to benefit a wide variety of long-range low-power IoT use cases.

\section*{Acknowledgments}
This was supported in part by National Science Foundation (NSF) Small Business Innovative Research (SBIR) Grant no. 2136427.

%Bibliography
\bibliographystyle{unsrt}  
\bibliography{references}

\end{document}